\documentclass[]{aastex631}

\pdfoutput=1

\shorttitle{Bayesian Synthesis in CB-SMBH}
\shortauthors{Kova{\v c}evi{\' c} et al.}
\usepackage{amsmath}
\usepackage{graphicx}
\usepackage{float}
\usepackage{multirow}
\usepackage{rotating}
\usepackage{bm}

\begin{document}

\title{Bayesian synthesis of  astrometric wobble and total light curves in \\
close binary supermassive black holes}

\correspondingauthor{Andjelka B. Kova{\v c}evi{\'c}}
\email{andjelka.kovacevic@matf.bg.ac.rs}

\author[0000-0001-5139-1978]{Andjelka B. Kova{\v c}evi{\'c}}
\affiliation{University of Belgrade-Faculty of Mathematics, Department of astronomy,
Studentski trg 16
Belgrade, Serbia}

\author[0000-0003-4042-7191]{Yu-Yang Songsheng}
\affiliation{Key Laboratory for Particle Astrophysics, Institute of High Energy Physics,\\ Chinese Academy of Sciences, 19B Yuquan Road, Beijing 100049, China}

\author[0000-0001-9449-9268]{Jian-Min Wang}
\affiliation{Key Laboratory for Particle Astrophysics, Institute of High Energy Physics,\\ Chinese Academy of Sciences, 19B Yuquan Road, Beijing 100049, China}

\author[0000-0003-2398-7664]{Luka {\v C}. Popovi{\'c}}
\affiliation{University of Belgrade-Faculty of Mathematics, Department of astronomy,
Studentski trg 16
Belgrade, Serbia}
\affiliation{Astronomical Observatory, Volgina 7, 11000 Belgrade, Serbia}

%from previous versions of AASTeX is now
%% depreciated in this version as it is no longer necessary. AASTeX 
%% automatically takes care of all commas and "and"s between authors names.

%% AASTeX 6.31 has the new \collaboration and \nocollaboration commands to
%% provide the collaboration status of a group of authors. These commands 
%% can be used either before or after the list of corresponding authors. The
%% argument for \collaboration is the collaboration identifier. Authors are
%% encouraged to surround collaboration identifiers with ()s. The 
%% \nocollaboration command takes no argument and exists to indicate that
%% the nearby authors are not part of surrounding collaborations.

%% Mark off the abstract in the ``abstract'' environment. 
\begin{abstract}

We test the potential of Bayesian synthesis of upcoming multi-instrument data  to extract orbital parameters and individual light curves of close binary supermassive black holes (CB-SMBH) with subparsec separations. Next generation (ng) interferometers, will {make possible} the observation of astrometric wobbles in CB-SMBH. Combining them with periodic variable  time-domain data from  surveys  like the Vera C. Rubin Legacy Survey of Space and Time (LSST), allows for a more information on CB-SMBH candidates compared to standalone observational methods.  
{Our method reliably determines binary parameters and  component fluxes from binary total flux across long-term, intermediate and short-term  binary dynamics and observational configurations, assuming ten annual observations,  even in short period "q-accrete" objects.}
{Expected} CB-SMBH astrometric wobbles {constructed} from  binary dynamical parameters, {might serve} in refining observational strategies for CB-SMBH. Combination of inferred mass ratio,  light curves of binary components, and observed photocenter wobbles can be a proxy for the activity states of CB-SMBH components.

\end{abstract}

%% Keywords should appear after the \end{abstract} command. 
%% The AAS Journals now uses Unified Astronomy Thesaurus concepts:
%% https://astrothesaurus.org
%% You will be asked to selected these concepts during the submission process
%% but this old "keyword" functionality is maintained in case authors want
%% to include these concepts in their preprints.

%% From the front matter, we move on to the body of the paper.
%% Sections are demarcated by \section and \subsection, respectively.
%% Observe the use of the LaTeX \label
%% command after the \subsection to give a symbolic KEY to the
%% subsection for cross-referencing in a \ref command.
%% You can use LaTeX's \ref and \label commands to keep track of
%% cross-references to sections, equations, tables, and figures.
%% That way, if you change the order of any elements, LaTeX will
%% automatically renumber them.
%%
%% We recommend that authors also use the natbib \citep
%% and \citet commands to identify citations.  The citations are
%% tied to the reference list via symbolic KEYs. The KEY corresponds
%% to the KEY in the \bibitem in the reference list below. 

\section{Introduction} \label{sec:intro}

Hierarchical formation theories propose hypothesis that close binary supermassive black holes (CB-SMBHs), characterized by subparsec mutual separations, could be formed in the centers of  galaxies as a consequence of galactic mergers \citep{1980Natur.287..307B, 1978MNRAS.183..341W}. These compact objects are predicted to emit gravitational waves (GWs) in the frequency range of $10^{-9}$ to $10^{-7}$ Hz \citep{2020ApJ...905L..34A,2019A&ARv..27....5B, 2004ApJ...611..623S}. The emission of gravitational launches the contraction of their orbits, leading to the anticipated  collision of the components in the system.

Earlier theoretical and simulation studies have developed a general model of CB-SMBHs. This model suggests a gaseous circumbinary disk encircling the CB-SMBH \citep[e.g.,][and references therein]{Bogdanovi__2011,10.1093/mnras/stt1787, dorazio15,2016A&A...588A.125R, 10.1093/mnras/stx1130, Bowen_2018}. Both black holes may accrete material from this disk, giving rise to two distinct mini-disks, each surrounding one of the black holes in the binary system. These mini-disks are assumed to be geometrically thin but optically thick accretion flows, portraying a scenario where electromagnetic and gravitational wave emissions could be detected well before the eventual merger and would continue until the culmination of the merger event \citep{2015ApJ...807..131S,2017ApJ...838...42B,2019ApJ...879...76B,10.1093/mnrasl/slu184, 10.1093/mnras/sty423, DEROSA2019101525, 2022LRR....25....3B}.

The capability to observe and characterize binary SMBHs across all phases of their evolutionary trajectory, extending from the wide separation \citep{2003ApJ...582L..15K} to more compact configurations \citep{2000SerAJ.162....1P,10.1093/mnras/stv1726, Charisi2016,Liu_2020, 2020ApJS..247....3S, 10.1093/mnras/staa1985, 10.1051/0004-6361/202039368,10.1093/mnras/staa2957, 2021MNRAS.505.5192P, 2022AN....34310073S}, would enable us to address the history, demography, origin, and co-evolution of SMBHs and galaxies throughout cosmic time \citep[e.g.,][]{2016IAUFM..29B.292K, 2021POBeo.100...29K}.

Distinctive characteristic of CB-SMBHs is that the thermal peak of their  emission is  correlated with $M_{tot}^{-0.25}$ \citep[see Figure 14 in][]{Guti_rrez_2022}. This property underscores a significant challenge in differentiating between binary and single active SMBHs, due to the expectation of abundant gas available for accretion in CB-SMBH systems \citep[e.g.,][]{2015ApJ...807..131S, Guti_rrez_2022}. The expected orbital periods of CB-SMBHs, being inversely proportional to the system’s total mass, span a range from a few days to centuries. This diversity in possible periodic  timescales necessitates the employment of a various observational strategies and instruments to identify and study CB-SMBHs across the wide ranges of masses and orbital periods \citep[see e.g.,][]{Guti_rrez_2022}.

Identification of CB-SMBH candidates hinges on indirect detection methods, owed to the inherent challenges in spatially resolving subparsec binaries \citep[see e.g., ][]{DEROSA2019101525,2022LRR....25....3B, 10.1086/511032,10.1088/0004-637X/725/1/249, pop12,10.1088/0004-637X/775/1/49,10.1088/0004-637X/789/2/140, Wang_2018, 10.3847/1538-4357/aaeff0,10.1051/0004-6361/202038733, 2021MNRAS.505.5192P, 2022AN....34310073S}. 
However, the most effective method remains the detection of distinct periodic modulations in the emission \citep{Guti_rrez_2022}.

At present, there is a noticeable trend towards the integration of various observational and data analysis techniques to detect binary SMBH. One example is the varstrometry technique \citep{2020ApJ...888...73H}, rooted in the concepts introduced by \citet{Shen_2012} and \citet{2015A&A...580A.133L}. This method uses the non-synchronous variation in flux, causing an astrometric shift in the combined photocenter measured at different observing epochs of dual and off-set quasars. This effect leads to photocenter variations at the $>$ mas level, which are detectable by GAIA astrometry \citep{2016A&A...595A...1G, 2021A&A...649A...1G, 2023ApJ...942...99G}.
Importantly, the combination of GRAVITY’s spectroastrometry \citep{10.1051/0004-6361/201730838} and reverberation mapping (RM) is showing promise for extracting three-dimensional insights into the emitting regions of CB-SMBH \citep[see SARM][]{2020NatAs...4..517W,10.1088/1674-4527/20/10/160,10.2298/SAJ2102001K}.

Furthermore, the dynamics of quasar variability, influenced by a plethora of factors including shocks in jets, dust clouds, accretion disk instabilities, proximal stellar activities, and CB-SMBH motion, can potentially disrupt the quasar photocenter's position and motion \citep{Popovic12, 2022A&A...660A..16S}. Therefore, photocenter perturbations may provide valuable information that can be instrumental in detecting  CB-SMBH systems.

%A rough estimate of the photocenter perturbation for an unresolved CB-SMBH with a period significantly shorter than the observational baseline can be approximated as a difference in the center of mass and photocenter so that astrometric wobble is $\propto \theta \frac{1-\delta}{1+\delta}$, where $\delta=q/l$ is the proportion of the secondary and primary component mass ($q=M_{2}/M_{1}$) and flux ($l=F_{2}/F_{1}$) ratios, whereas $\theta$ is the size of the orbit projected on the sky. Then we estimated that expected  photocenter  wobble is $\sim 10 \mu$as,  assuming an astrometric orbit size of \(34.2 \mu\text{as}\) at a redshift of $\sim 0.1697$  \citep[as estimated for the best candidates for observations by GRAVITY+, see][]{Dexter_2020}, given that that $q \sim 0.5$ and $ l \sim 0.27$.

\begin{figure*}
{\includegraphics[width=0.33\textwidth]{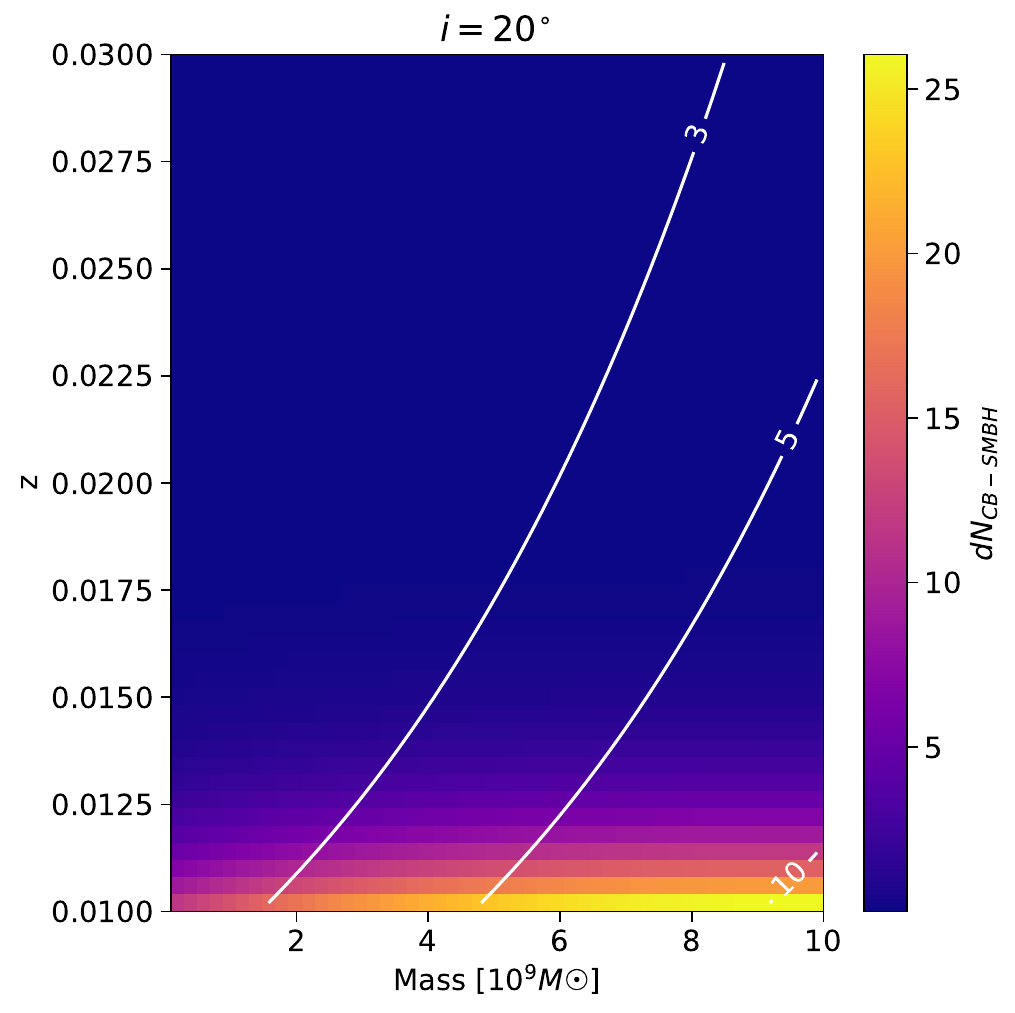}} 
{\includegraphics[width=0.33\textwidth]{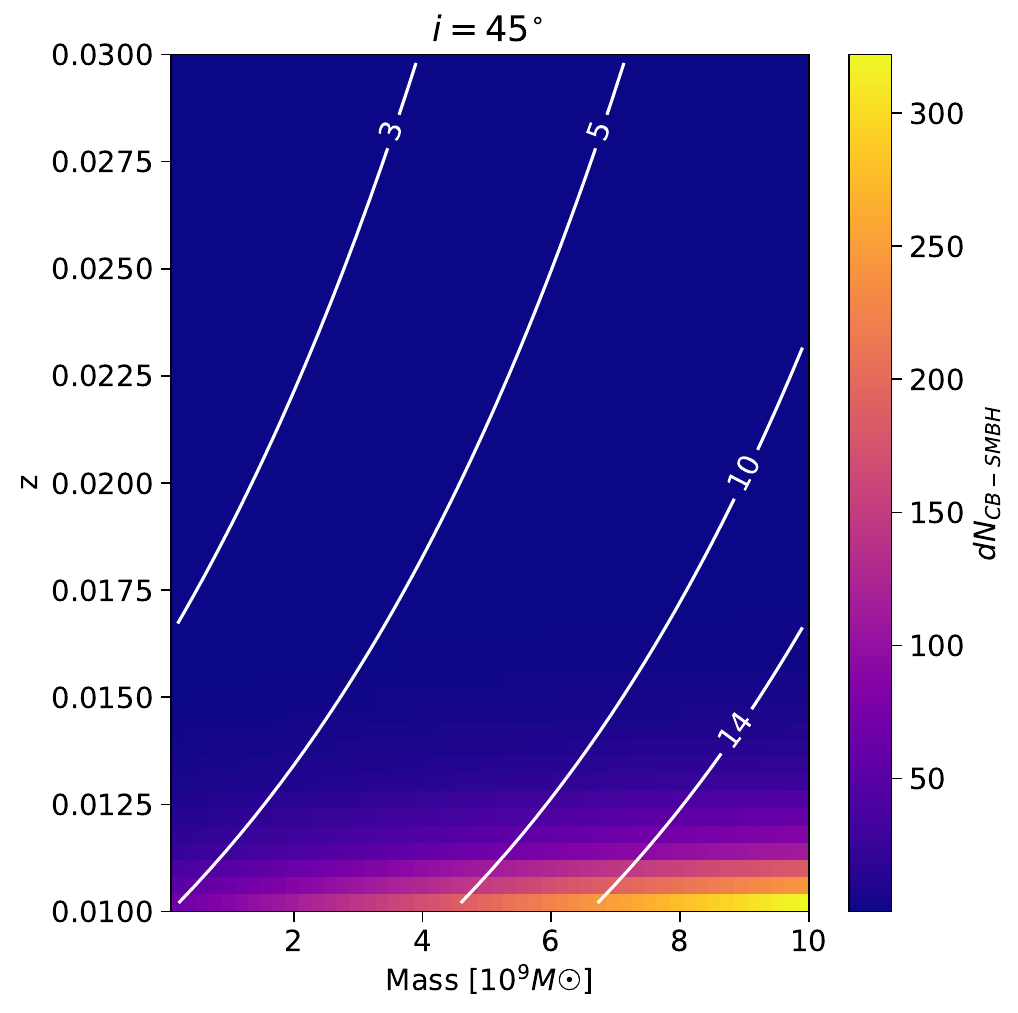}} 
{\includegraphics[width=0.33\textwidth]{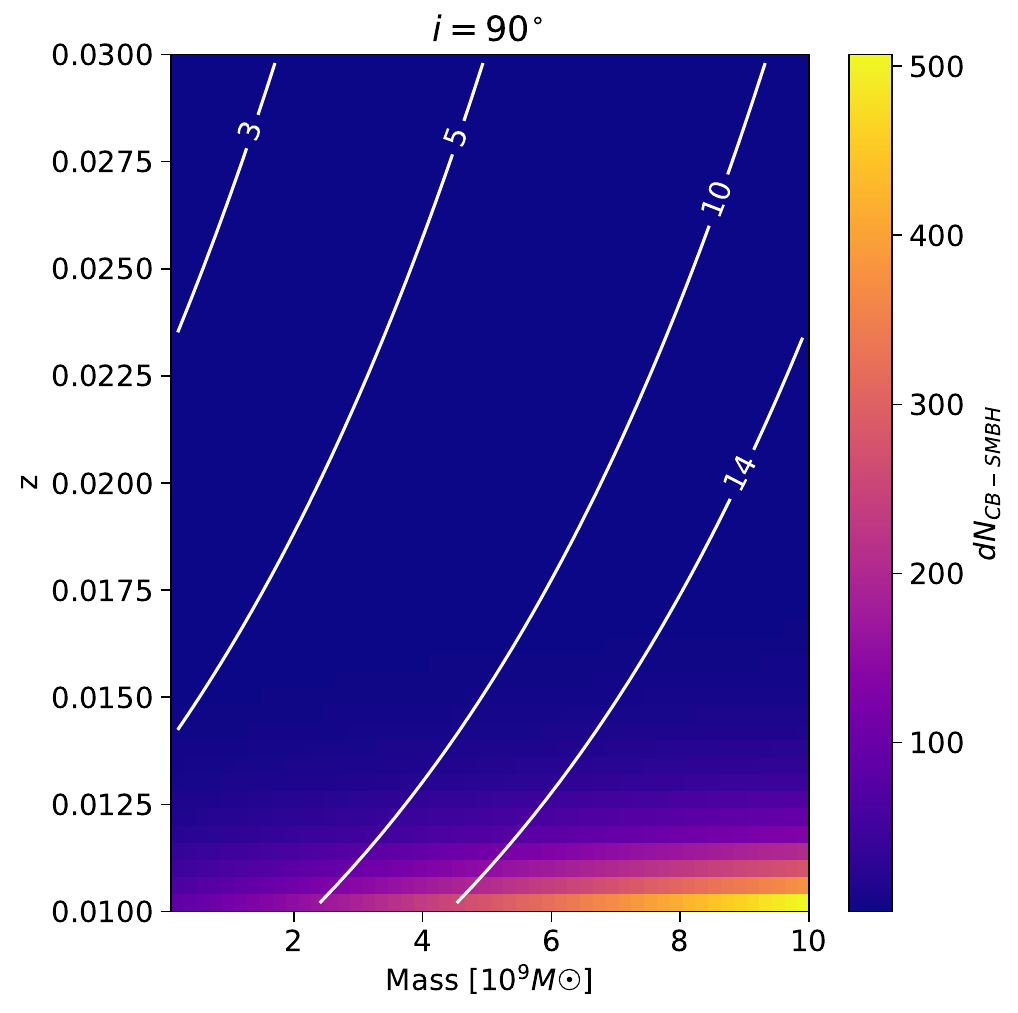}} 
\caption{The estimated number of detectable CB-SMBHs ($d N_\mathrm{CB-SMBH}$) with periods $P\in [0.5,1]$ year as a function of the system's total mass  and redshift. 
 The  binary systems   inclinations
 are $20^{\circ}$ (left) $\geq 45^{\circ}$ (middle) and $90^{\circ}$. Mass and luminosity ratios are considered within the ranges 
$q=[0.1,0.67]$ and $l=[0.1,1]$, respectively. White contour lines correspond to constant expected astrometric wobble values in $\mu$as, indicating the detectability threshold for such systems.
}
\label{fig:population}
\end{figure*}

The next-generation Event Horizon Telescope \citep[ng EHT][]{galaxies11030061} with advanced direct radio interferometry imaging and the GRAVITY+ instrument onboard the VLT \citep{mpe.mpg.de/7480772/GRAVITYplus_WhitePaper.pdf} with astrometry using IR interferometry can probe photocenter perturbation. Employing "super-resolution" techniques, the ngEHT's baseline angular resolution of $ 15 \mu $as can be substantially improved,  potentially resolving features as small as a $\sim$ few $ \mu $as in size \citep{Broderick_2020, galaxies11030061}. 
Observations with the GRAVITY instrument have shown its capability to reach an angular resolution of $10\mu$as, in the K band, but the target luminosity and substantial exposure times, often exceeding several hours, are necessary \citep{2022A&A...665A..75G, Dexter_2020,Li_2023}. Although GAIA's astrometric precision allows for the detection of $10 \mu$as centroid shifts in luminous sources, it cannot directly image CB-SMBHs \citep{PhysRevD.100.103016}. Given these instruments' characteristics and the predicted photocenter wobble for subannual CB-SMBH systems as illustrated in Figure \ref{fig:population}, there is a robust potential for  detecting subtle signals of astrometric wobble  with the  next generation interferometers.

Similarly, as the varstrometry approach is applicable to AGN dual systems with large mutual separation and relies on back-and-forth photocenter motion on the line linking the two components \citep{2020ApJ...888...73H}, we suggest a generalized method for CB-SMBH binaries with non-fixed geometry. A similar concept, known as "variability induced motion," was applied to unresolved star binaries \citep{1996A&A...314..679W}. The technique we propose compensates for the photocenter motion \citep[see also][]{PhysRevD.100.103016} on a Keplerian orbit with regard to the center of mass. Unlike varstrometry, we cannot anticipate a perfect correlation between the instantaneous photocenter shift and the photometric flux due to binary motion, which imposes relying on Bayesian inference.
Moreover, given that the joint variability light curve of CB-SMBH stores both the stochastic and periodic variability of the system \citep[e.g.][]{10.1093/mnras/stab1856}, our method also decomposes the light curve into its separate components using Gaussian processes. Targeted CB-SMBH candidates by ngEHT and GRAVITY+ \citep{doi.org/10.3847/1538-4357/ab3c5e} should be brighter in the near-IR \citep{DOrazio_2018, mpe.mpg.de/7480772/GRAVITYplus_WhitePaper.pdf} and hence accessible via all-sky time-domain surveys in the near-IR \citep{DOrazio_2018}, e.g., such as the Vera C. Rubin Large Synoptic Survey Telescope \citep{2019ApJ...873..111I, 2022ApJS..258....1B}.
Importantly, our method capitalizes on the synergy of information from both interferometric and time-domain data, enhancing our ability to accurately interpret the observations of upcoming multiband surveys.

The structure of the article  is laid out as follows. In Section \ref{sec:model}, we detail the main principles of our method, and provide two tiered Bayesian protocol for  simulations of astrometric wobble and time domain observations  of CB-SMBH, alongside the subsequent recovery of binary parameters and individual light curves.  We present our results in Section \ref{sec:result}.
A discussion on the  broader implications of our results can be found in Section  \ref{sec:discussion}.  We summarize  our study with conclusions in Section \ref{sec:concl}.

\section{Method}\label{sec:model}

In this section, we provide an analytical description of the astrometric wobble of CB-SMBH system with non-fixed geometry. Then, we introduce a Bayesian protocol \citep[for further details, see][]{2022A&A...663A..99K} designed to extract both the binary's orbital parameters and the individual emission light curves of its constituent components. Notably, our method  can be universally applied to analyze any dataset comprising the astrometric wobble and the total emission light curve of a quasar, making it instrumental agnostic.

\subsection{Analytical Framework for Astrometric Wobble in CB-SMBH Systems with  Orbital Motion}\label{sec:prva}

In the case of an unresolved binary system where the orbital motion of the components is significant, it is essential to consider the photocenter perturbation caused by this motion. This perturbation,  ($\bm{\delta}_{ph}$) approximated as the difference between the photocenter and the center of mass \citep[see also][]{10.1093/mnras/staa1522}. Consequently, in CB-SMBH systems the perturbation vector $\bm{\delta}_{ph}=(\tilde{x}(t), \tilde{y}(t))$ \citep{2012asas.book.....V} in CB-SMBH,  can be approximated as follows:

\begin{align}
\begin{split}
\bm{\delta}_{ph}&\propto\frac{M_{1}\bm{\xi}_{1}+M_{2}\bm{\xi}_{2}}{M_{1}+M_{2}}-\frac{F_{1}\bm{\xi}_{1}+F_{2}\bm{\xi}_{2}}{F_{1}+F_{2}}\\
    &=(\mu-l(t))(\bm{\xi}_{2}-\bm{\xi}_{1})
\end{split}
    \label{eq:photonew}
\end{align}
\noindent where $(M_{i}, F_{i}), i=1,2$ denote the masses and fluxes of the components in the binary. The mass and flux ratios are given by $q=M_{2}/M_{1}$ and $f(t)=F_{2}/F_{1}$, respectivley. The reduced  mass and flux  are represented as  $ \mu=\frac{M_{2}}{M_{1}+M_{2}}$ and $l(t)=\frac{F_{2}}{F_{1}+F_{2}}$, respectively. The vectors $\bm{\xi}_{i}=(x_{i},y_{i}), i=1,2$ represent the projected positions of the binary  components on the sky.\footnote{Further details on  the binary components position  projection onto the sky plane, can be found in \citet{2022A&A...663A..99K}.} 

In coordinates, the perturbation vector translates to:

\begin{align}
  \bm{\delta}_{ph} &\propto \left\{
  \begin{array}{l}
      \tilde{x}(t) = -(\mu-l(t))s_{x}(t) \\
      \tilde{y}(t) = -(\mu-l(t))s_{y}(t)
  \end{array}
  \right. \\
  \label{eq:photonew1}
\end{align}

\noindent where, $\bm{s}(t)=(s_{x}(t)=x_{2}(t)-x_{1}(t), s_{y}(t)=y_{2}(t)-y_{1}(t))$. A detailed explanation of this expression is provided in the Appendix  \ref{sec:apend1} (see Equation \ref{eq:compose}), and further context can be found in \citet{1996A&A...314..679W}. 

The equations are rewritten in a more detailed form below (for further details, see Appendix \ref{sec:astrom}):

\begin{align}
\begin{split}
\tilde{x}(t) &=-(\mu-l(t))s_{x}(t) \\
&=-(\mu-l(t))a \Bigg[\left[\sin \Omega \cos(\omega) + \mathcal{I}\cos{\Omega} \sin(\omega)\right] \left[\cos E(t) - e\right] \\
&\quad + \left[-\sin \Omega \sin(\omega) + \mathcal{I}\cos{\Omega}\cos(\omega)\right] \Bigg]\sqrt{1-e^2}\sin E(t),\\
\tilde{y}(t) &=-(\mu-l(t))s_{y}(t)  \\
&=-(\mu-l(t))a \Bigg[\left[\cos \Omega \cos(\omega) - \mathcal{I}\sin\Omega \sin(\omega)\right] \left[\cos E(t) - e\right] \\
&\quad + \left[-\cos \Omega \sin(\omega) - \mathcal{I}\sin\Omega\cos(\omega)\right]\Bigg] \sqrt{1-e^2}\sin E(t).
\end{split}
\end{align}

\noindent where $a$ denotes the semimajor axis of the secondary, $e$ is the eccentricity of the orbit, $E(t)$ is the eccentric anomaly, and $\Omega$, $\omega$, and $\mathcal{I}=\cos i$ are the Euler orbital angles as defined in Appendix \ref{sec:astrom}. The Euler angles $\Omega$ and $\omega$, along with the orbital inclination $i$, play a crucial role in calculating the astrometric wobble as they influence the orientation of the binary orbit in space. Moreover, the quantity $(\mu-l(t)){a}$, roughly corresponds to the semimajor axis of the photocenter orbit due to the Keplerian motion of the binary. 
These relations allows us to connect the physical parameters of the binary system (such as masses, fluxes, and orbital parameters) to the observable astrometric signatures, providing valuable insights into the dynamics of the system. 
In this formalism, the mass and light ratios ($\mu$ and $l(t)$) are important in shaping the astrometric wobble. 
An equivalence in the values of mass and luminosity ratios ($\mu \approx l(t)$) minimizes the astrometric wobble. In contrast, a pronounced disparity in these values ($\mu \neq l(t)$) amplifies the wobble. 

The pioneering observations by the EHT  of the shadow of a SMBH) with mass $ (6.5 \pm 0.7) \times 10^9 M_{\odot}$ at the center of M87 \citep{Akiyama_2019} have spurred interest in the search for binary companions within such massive systems. \citet{2019MNRAS.488L..90S} proposed that M87, situated in a region known for frequent galactic mergers \citep{Hopkins_2006}, may host a binary companion. Prior to this, PSO $J334.2028+01.4075$, a radio-loud quasar at $z = 2.060$, was identified as a strong candidate for a CB-SMBH, supported by extensive photometric observations \citep{2015ApJ...803L..16L}. Its periodic variability of 
 $ 542 \pm 15  $ days and the corresponding estimated black hole mass of $\log(M_{\text{tot}}/M_{\odot}) = 9.97 \pm 0.50 $ suggest an orbital separation of approximately $\sim 0.006^{+0.007}_{-0.003}$ pc. This separation is indicative of a system that may be approaching the GW-driven orbital decay phase, located at the epoch of peak SMBH mergers. This underscores the capabilities of current surveys and foreshadows the  potential of the upcoming LSST in the discovery and characterization of CB-SMBHs.

Informed by these findings, we are encouraged to extend our exploration to binary CB-SMBHs with subannual orbital periods, within the upper mass range $\sim 10^{10} M_{\odot}$ \citep{2019AJ....157..148G,2019ApJ...887..195M,2017MNRAS.471.2321D,2009MNRAS.399L..24G,2020MNRAS.496.2309O,2015ApJ...799..189Z,
10.1093/mnras/stad587, 2014MNRAS.442.2809W,1998AJ....115.2285M,2020ApJ...899...76J,2016A&A...587A..43S,2012ApJ...756..179M,
2012MNRAS.427...77V,2016Natur.532..340T,2023ApJ...950...68E,2015ApJ...808...79F,2015MNRAS.450L..34G,2022ApJS..261....5M,
2015ApJ...803L..16L,2013ApJ...777..163H,2016MNRAS.456..538Y}.

{The Figure  \ref{fig:population} visualizes (see Section  \ref{sec:apendno}) the estimated detectability CB-SMBHs with   $P \rightarrow 1$  year, a range that is recognized as a 'binary candidate desert' 
\citep[see][and their Figure 12]{Charisi2016}. We find that detectability of such systems is maximized for systems with greater total mass, lower redshifts, and orbits with higher inclinations, as these conditions yield stronger astrometric signals. }
Overlaid contour lines correspond to constant astrometric wobbles, establishing benchmarks relative to the detection capabilities of the astrometric instruments. The ngEHT is sensitive to wobbles  $\geq 3\mu\mathrm{as}$, while GRAVITY+ threshold lies $\geq 10\mu\mathrm{as}$, which is also within the expected performance envelope of GAIA. It is important to note that astrometric wobbles $\sim 2\mu\mathrm{as}$) are at the threshold of ngEHT detectability, indicating that objects producing such  signals are on the cusp of observability. Given that the expected CB-SMBH orbital periods at this GW emission stage are months to years, these objects are accessible via multi-epoch observations with the ngEHT \citep{DOrazio_2018}.
As for the GRAVITY+ candidates, CB-SMBH candidates with single, offset, moving broad emission lines \citep{Dexter_2020} are mostly observed at $z \sim 0.2$, with apparent magnitudes of $V\lesssim 18$ and $K\lesssim 15$ \citep{2017MNRAS.468.1683R,2019MNRAS.482.3288G}. {In our simulations, we sampled a binary configuration from this specific parameter space with orbital period of  $\leq 1$ year, showcasing the capabilities of future synergies between astrometry and time-domain surveys, for identifying CB-SMBHs in regions once overlooked, thereby contributing to a more comprehensive understanding of binary supermassive black hole populations. However, we also examine one configuration with period 1.3 year, to illustrate the dynamics of systems with periods marginally exceeding our primary interval, shedding light on their characteristics as they converge towards 1 year from above.}

\subsection{Bayesian Protocol for Simulating and Synthetizing CB-SMBH Multi-instrument  Data}\label{sec:Bayesprotocol}

We outline a two-tier Bayesian protocol \citep[see also][]{2022A&A...663A..99K}. The first tier involves a Simulator that  generates observational multi-instrument data encompassing the astrometric wobble and optical light curves, predicated upon specified characteristics of the CB-SMBH. The second tier, the Bayesian Solver, processes this simulated data to extract 'solutions' that reveal the decomposed flux of, for instance, the secondary component, along with the orbital parameters of the CB-SMBH.

\subsubsection{First tier:  Simulator}\label{sec:simulator}

The CB-SMBH observed astrometric photocenter and flux data are simulated with procedures  given in Appendix \ref{sec:apend1}. The observed total light curve of CB-SMBH includes the  Doppler boosting  (Section \ref{sec:doppler}) and stochastic variability (Section \ref{sec:stochastic}).
We are considering higher line of sight angles, so that detected
radiation of CB-SMBH is anticipated to be strongly modulated by Doppler boosting. We note that for face on objects, gravitational redshift and traverse Doppler effects could be more important \citep{Guti_rrez_2022}. 
  For stochastic variability, we adopted  model describing the stochastic emission $F_{i}$ of each minidisk in the context of the lamp-post model, thin-disk geometry, and damped random walk (DRW) for the driving function, where the transfer function is assymetric \citep{2020A&A...636A..52C,2022ApJS..262...49K}. Anticipations of such  light curves are substantiated by forthcoming extensive time domain surveys like the LSST  \citep{2020A&A...636A..52C,2022ApJS..262...49K}.
  The errors $\epsilon_{astfl}$ for each observed point in the  astrometry and time domain data are independent and identically distributed such as that  their probability density functions (PDF’s) are normal with mean zero and standard deviations between 3\% and 10\% of the data \citep[e.g.,][]{Dexter_2020, 2019ApJ...873..111I}.
  To avoid using exactly the same
model to generate the observed data and to find the inverse solution
 an additional jitter was included in the model when generating the data \citep[see][]{2009A&A...494..769T}.

Regarding the astrometric tracking, the capabilities are advancing to a point where precision can reach $1 \mu$ as for ng EHT \citep[see][and references therein]{DOrazio_2018}. With this enhanced capability in mind, we consider binaries of $M_{tot}\geq 10^{8}M_{\odot}$ with a physical size of $a\leq 0.1$ pc at a distance of $D\sim 70$Mpc.
For each black hole in the system, the radii of the inner ($r_{0i}$, $i=1,2$) and outer edges ($r_{\text{out}i}$, $i=1,2$) of the accretion disks are computed under the assumption of non-spinning black holes ( Section~\ref{sec:stochastic}):
$r_{01}=0.0056 \, \text{ld},  r_{02}=0.0037 \, \text{ld} 
, r_{\text{out}1}=0.0092 \, \text{ld},  r_{\text{out}2}=0.0047 \, \text{ld} $.

The  sum of outer radii  $\sim 0.014$ ld is notably less---by over three orders of magnitude---than the pericenter distance of $5$ ld. This ensures that even during the closest approach, the accretion disks are not subject to disruptive tidal forces \citep[see e.g.,][]{Wang_2018,10.1093/mnras/stac3266}, thus substantiating stability of the accretion disks within the binary system.

%The parameters listed in the second section of Table \ref{tab:campaigns} are adopted based on typical values %observed in quasars \citep[see e.g.,][and reference therein]{Kovacevic2021,Kovacevicc2022}. 
%We selected a specific set of parameters ($\Omega = 180^\circ$, $\omega = 30^\circ$, $LOS = 20^\circ$) to simulate %observational campaigns of a binary system, showcasing a particular orbital configuration where the ascending node %opposes the reference direction, and the pericenter lies $30^\circ$ counterclockwise from the ascending node. 
%This configuration projects an elliptical orbit with its semi-major axis angled at $30^\circ$ to the line of sight %and rotated $180^\circ$ in the sky plane, featuring counterclockwise orbital motion. Despite this specific %scenario, we emphasize the versatility of our developed method. It is adept at extracting binary parameters and %the secondary component flux from photocenter motion and total flux, irrespective of the binary's particular %parameter set, affirming its universal applicability.
{We strategically selected binary configurations $Ci, i=1,6$ to encompass a range of dynamical behaviors and observational cycles ($L_{oc}$), which represent the observational time baseline relative to the orbital period of the binary system. Alongside $L_{oc}$, constant astrometric cadence density of 10 observations per year, is informed by the practical constraints and opportunities presented by current and forthcoming observational facilities \citep[see][]{Fang_2022}. Despite the potential for acquiring a photometric light curves with very dense LSST cadences, we opt for the adversial photometric cadence  identical to astrometry  to illustrate our method's efficacy  in data-sparse scenarios.  Configurations $C1-C5$
 account that accretion rate ratio of binary components satisfy 
 $\eta=\frac{\dot{M}_{2}}{\dot{M}_{1}}=(1+0.9q)^{-1}$ \citep{2020ApJ...901...25D}.
Configuration $C6$ focuses on the dynamics of "q-accrete" as an alternative where the accretion rate of each component  is scaled proportionally to its mass, i.e., the fluxes of the binary quasar system are expected to be directly proportional to the masses of the individual black holes \citep{10.1093/mnras/stz150}. In this case, we adopt $q=0.67$  \citep{Farris_2014}, since the orbital frequency of the binary is still significant \citep[see Figure 9 in][]{Farris_2014} and the
luminosity of the primary mini-disc may also become significant \citep[see Figure 8 in][]{10.1093/mnras/stab3713}.  }
\begin{table*}
    \caption{Binary configurations  with a constant density of 10 observations per year in astrometry and light curves.}
    \label{tab:parameter_combinations}
    \centering
    \setlength{\tabcolsep}{4pt} % Adjust column spacing for better readability
    \begin{tabular}{ccccccccc}
    \hline
    \hline
  Configuration& $e$ & q&$i$ [$^\circ$] & $\Omega$ [$^\circ$] & $\omega$ [$^\circ$] & $M$ [$M_\odot$] & $P$ [years] & $L_{oc}$ \\
    \hline
    C1 & 0.3 & 0.25 & 60 & 45 & 60 & $10^9$ &2.28& 2 \\
    C2 & 0.5 & 0.4 & 30 & 45 & 60 & $10^9$ &1& 1 \\
    C3 & 0.1 & 0.25 & 30 & 205 & 100 & $10^9$ &9& 2/3 \\
    C4 & 0.7 & 0.3 & 60 & 105 & 200 & $10^9$ &3.01& 1/2 \\
    C5 & 0.7 & 0.3 & 60 & 105 & 200 & $10^8$ & 2.5&1/2 \\
    C6 & 0.7 & 0.67 & 60 & 105 & 200 & $10^{10}$ &1.3& 1.5 \\
    \hline
       \multicolumn{8}{c}{General parameters} \\
    \hline     
    flux and astrometry error model &$\sigma_{astfl}$&\multicolumn{7}{c} {Normal$(0,s), s\sim$Uniform$(0.03,0.1)$}\\
Wavelength of observed light curve [\AA] & $\lambda$ & \multicolumn{7}{c} {$3670.7^{\mathrm{a}}$} \\
Radiative efficiency of quasar & $\epsilon$ & \multicolumn{7}{c} {$0.1$} \\
Distance to quasar [Mpc]& $D$ & \multicolumn{7}{c} {$70$Mpc} \\
DRW characteristic amplitude $[mag\, day^{-1/2}]$ &$\tilde{\sigma}$& \multicolumn{7}{c} {Eq. 22 in \text{\cite{Kelly_2009}}} \\
DRW time scale[day]&$\tau_{DRW}$& \multicolumn{7}{c} {Eq. 25 in \text{\cite{Kelly2009}}} \\
DRW mean {flux density} of quasar light curve [$\mathrm{erg\,cm}^{-2}\,\mathrm{s}^{-1}\,\mathrm{Hz}^{-1}$]& $m_{0}^{\mathrm{b}}$& \multicolumn{7}{c}{} \\
\hline
\hline
\multicolumn{8}{l}{$^{\mathrm a}$ The LSST \texttt{u}-band filter.}\\
\multicolumn{8}{l}{%
    \parbox{\textwidth}{%
        $^{\mathrm b}${ $m_{0}$ is derived from the calculated luminosity of the quasar, which is based on its black hole mass. This luminosity is then converted into a flux density observed at Earth, taking into account the distance of the quasar and the wavelength of the LSST \texttt{u}-band filter.}
    }%
}\\
 \end{tabular}
\end{table*}

\subsubsection{Second tier: Solver}\label{sec:Bayes}

As already mentioned we use the  Solver set of equations  for fitting procedure, but without jitter term. Denoting the astrometric observations of photocenter motion by $\Theta^{o}$, the flux observations  by $F^{o}_{tot}$, then the posterior
distribution of the orbital parameters of binary motion and flux of one of components factored  out of  total flux  of binary ($u=(a,e,P,i,\omega,\Omega, F_{j}), j=1,2 $) via Gaussian process
 is:
\begin{equation}
\begin{split}
P(u|\Theta^{o}, F^{o}_{tot})& \propto P(\Theta^{o}, F^{o}_{tot}|u)\cdot P(\theta, \alpha_{GP})= \\
&P(\Theta^{o}|\theta) \cdot P(F_{j}|F_{1}+F_{2},\alpha_{GP}) \cdot P(\theta, \alpha_{GP}), j=1,2
\end{split}
\end{equation}
\noindent The probability density functions contributing to the equation above could be obtained as follows. The function $P(\Theta^{o}|\theta)$ is the log-likelihood of fitting the astrometry data:

\begin{equation}
\mathcal{L} = -\frac{1}{2} \sum_{i=1}^{N} \left[ \frac{(x_i - x_{\text{model}}(\theta))^2}{\sigma_{x_i}^2} + \frac{(y_i - y_{\text{model}}(\theta))^2}{\sigma_{y_i}^2} \right]
\label{eq:posteriortotal}
\end{equation}

\noindent  where  $N$ is the number of observed astrometric data points,$x_i$ and $y_i$ are the observed positions of the photocenter,
$x_{\text{model}}(\theta)$ and $y_{\text{model}}(\theta)$ are the model-predicted positions of the photocenter using model parameters $\theta$,  and  uncertainties of observations are $\sigma_{x_i}$ and $\sigma_{y_i}$.
Next, a detailed derivation of $P(F_{i}|F_{1}+F_{2},\alpha_{GP})$ is provided in Section \ref{sec:sectionB0} (see Eq. \ref{eq:loglikegp1}). This probability density represents the log-likelihood of decomposing $F_{i}$ from the total flux of a binary system using a Gaussian process with parameters $\alpha_{GP}$. The prior probability density $P(\theta, \alpha_{GP})=P(\theta)P(\alpha_{GP})$ is a simple product of priors on orbital and GP parameters (see Table \ref{tab:priors}).
Then the posterior distribution  $P(u|\Theta^{o}, F^{o}_{tot})$, as given by Eq \ref{eq:posteriortotal}, is explored through  Bayesian framework in Python \texttt{PyMC} package \citep{Salvatier2016, pymc_2022}.  This program samples the posterior probability of the model
given the data, including prior probabilities for the parameters.

{The priors for the orbital parameters are sampled from distributions as given in the Table \ref{tab:priors}, with parameters  based on inspection of both the total light curve and astrometric observation. For example, in the case of C1, the  total light curve suggests a relatively stable and symmetric flux behavior, indicating a lower range for eccentricity ($e$) and a moderate range for orbital period ($P$). Astrometric observations reveal consistent positional changes indicative of a well-defined orbital plane, constraining parameters such as inclination ($i$).
The priors for configuration C1 include a uniform distribution  $e=\texttt{Uniform}(0.1, 0.4)$ and a normal distribution for $P$ centered around 2 years. In addition, priors for $i$ and $\Omega$ may be constrained to narrower ranges to reflect the stable orbital plane inferred from astrometric data.
For the configuration C3, the total light curve exhibits significant flux variations and irregularities, suggesting a wider range for eccentricity ($e$) and a longer orbital period ($P$) to accommodate the observed dynamical behavior.
Astrometric measurements reveal complex positional changes, so that  a broader range $(0, 2\pi)$ for $\omega$ and $\Omega$ priors  is used.
The priors for $e$  span  $\texttt{Uniform}(0.0.01, 0.6)$ while $P$ could follow a normal distribution centered around 7 years. Additionally, priors for $\omega$ and $\Omega$  are  $\texttt{Uniform}(0, 2\pi)$ to capture the complexity observed in the astrometric data.}
The error priors are drawn from normal distributions {that are} centered at expected errors of artificial data {and have a} standard deviation of $5\%$.

For the GP model, we did not employ DRW for recovering light curves. The Bayesian framework allows the use of any type of GP kernels. While there is a plethora of GP kernels that can be applied, in this study, we specifically used the squared exponential kernel (SE), which has been tested by   \citet{2017Ap&SS.362...31K}:

\begin{align}
    k=\alpha_{SE}^{2} \exp\Bigg(\frac{-(t_{i}-t_{j})^{2}}{2l_{SE}^{2}}\Bigg)
\end{align}
where $\alpha_{SE}$ sets the amplitude of the Gaussian process and has the same units as the flux, and $l_{SE}$ sets the length scale of the Gaussian process and has units of time $t$.
The Bayesian framework provides a natural mechanism to determine the most probable values of $\alpha_{SE}$ and $l$ through sampling priors of their ranges and optimization of Gaussian process models.
{For unfavorable configurations, more structured GP kernels could be used such as: 
$$k(X,X') =  \left(\alpha_{\text{SE}}^2 \exp\left(-\frac{(t_i - t_j)^2}{2l_{\text{SE}}^2}\right) + \text{diag}\left(D(X,\theta)^{(3+\alpha_{D})}\right)\right)$$
where covariance matrix of Doppler boost for binary parameters $\theta$ is given by $\text{diag}\left(D(X,\theta)^{(3+\alpha_{D})}\right)$.}

\begin{table*}
	\caption{Priors for the Bayesian solver for flux decomposition and orbital elements estimate.
}
	\centering
	\begin{tabular}{lcc}
		\hline\hline
		Description & Variable&Distribution \\
		\hline
		 \multicolumn{3}{c}{orbital parameters}\\
		 mass ratio of components&$q=M_{2}/M_{1}$ & Uniform  \\
		  total mass&$M_{tot}$ & Normal  \\
		 	eccentricity&$e$ & Uniform \\
		orbital period&	$P$&Normal   \\
		  argument of pericenter&	$\omega$&Uniform   \\
		 	
longitude of the ascending node&	$\Omega$&Uniform   \\
		 \multicolumn{3}{c}{GP kernels parameters$^{\mathrm a}$}\\
{amplitude of GP kernel for} $F_1$&$	\alpha_{1}$ & Normal$((1/(1+\eta))<F_{tot}>, 0.05)$ \\
amplitude of GP kernel for $F_2$&$	\alpha_{2}$ & Normal$((\eta/(1+\eta))<F_{tot}>, 0.05)$  \\

			\hline
				\multicolumn{3}{l}{$^{\mathrm a}$ kernels lenghtscales $l=\Gamma(\alpha=2, \beta=1/10)$. } \\
			\multicolumn{3}{l}{\small $^{\mathrm b}<F_{tot}>$ is average of observed total flux of CB-SMBH.} \\
	\end{tabular}
	\label{tab:priors}
\end{table*}

\section{Results }\label{sec:result}

\begin{figure}[p]
    \centering
    % First image and subcaption
    \begin{minipage}{0.98\textwidth}
        \centering
        \includegraphics[width=0.98\textwidth]{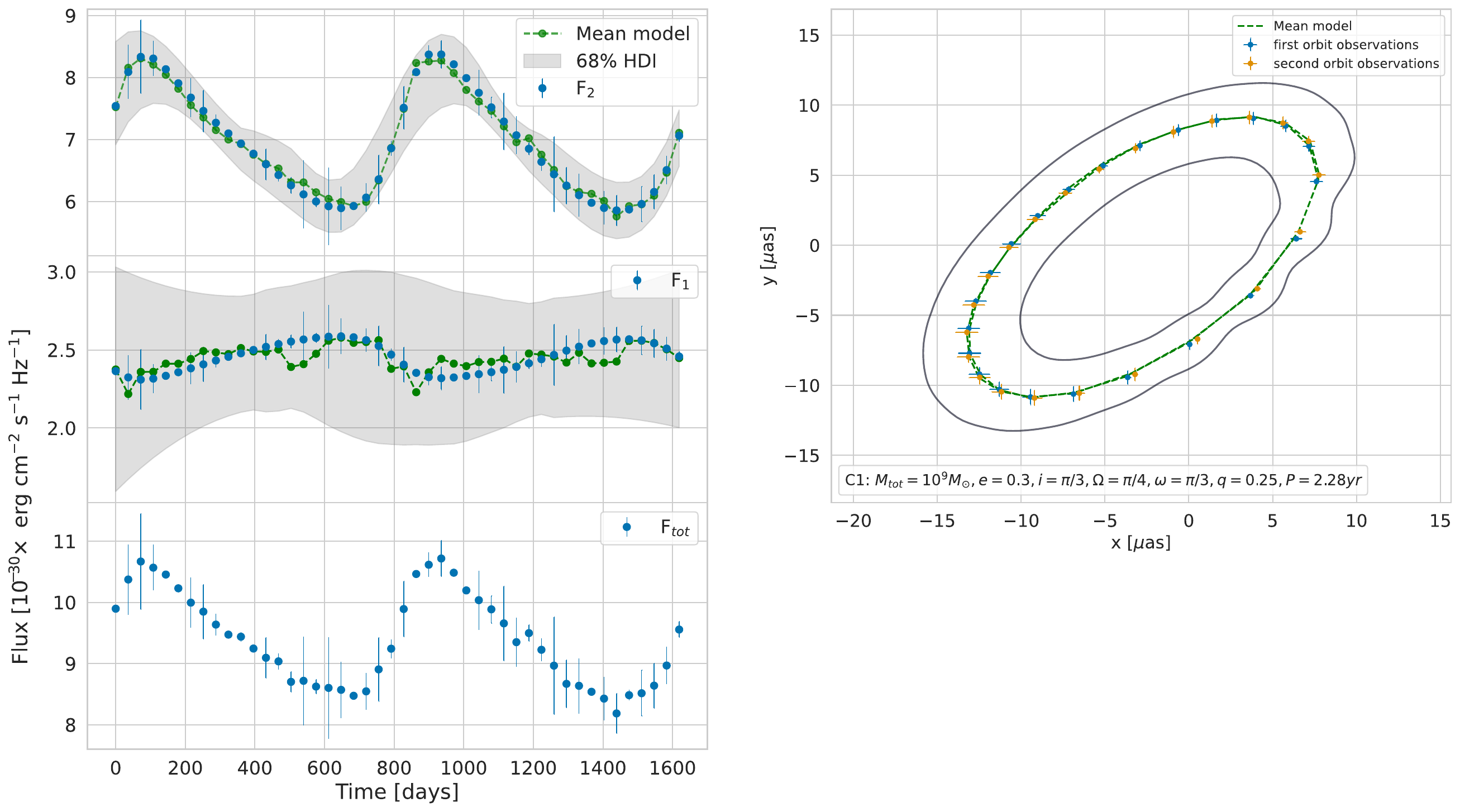}\\
        {(a) $C1$}
    \end{minipage}
    
    % Second image and subcaption
    \begin{minipage}{\textwidth}
        \centering
        \includegraphics[width=\textwidth]{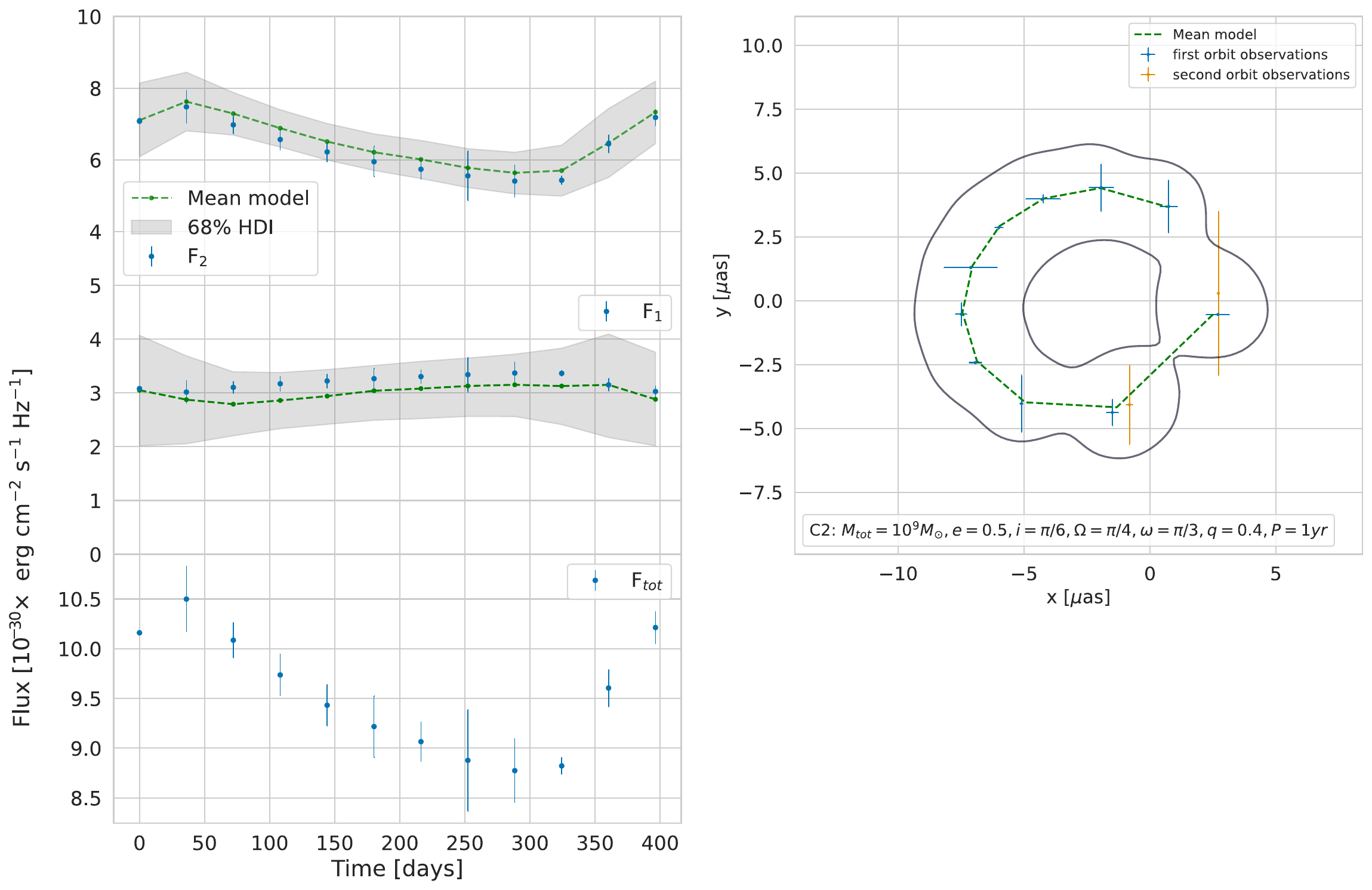}\\
        {(b) $C2$}
    \end{minipage}
    \caption{Bayesian inference of binary system dynamics through disentangled fluxes and astrometric wobble for  configurations (a) $C1$ and (b) $C2$. Top to bottom, the left panel shows disentangled secondary ($F_2$) and primary ($F_1$) fluxes, followed by the binary total observed flux ($F_{tot}$). Observations are marked by error bars, dashed green lines are the mean models, while grey shading indicates the 68\% HDI. On the right, astrometric wobble observations are coupled with the mean model, highlighted by 68\% HDI contour lines.}
    \label{fig:Fig1}
\end{figure}

\begin{figure}[p]
    \centering
    % First image and subcaption
    \begin{minipage}{0.98\textwidth}
        \centering
        \includegraphics[width=0.98\textwidth]{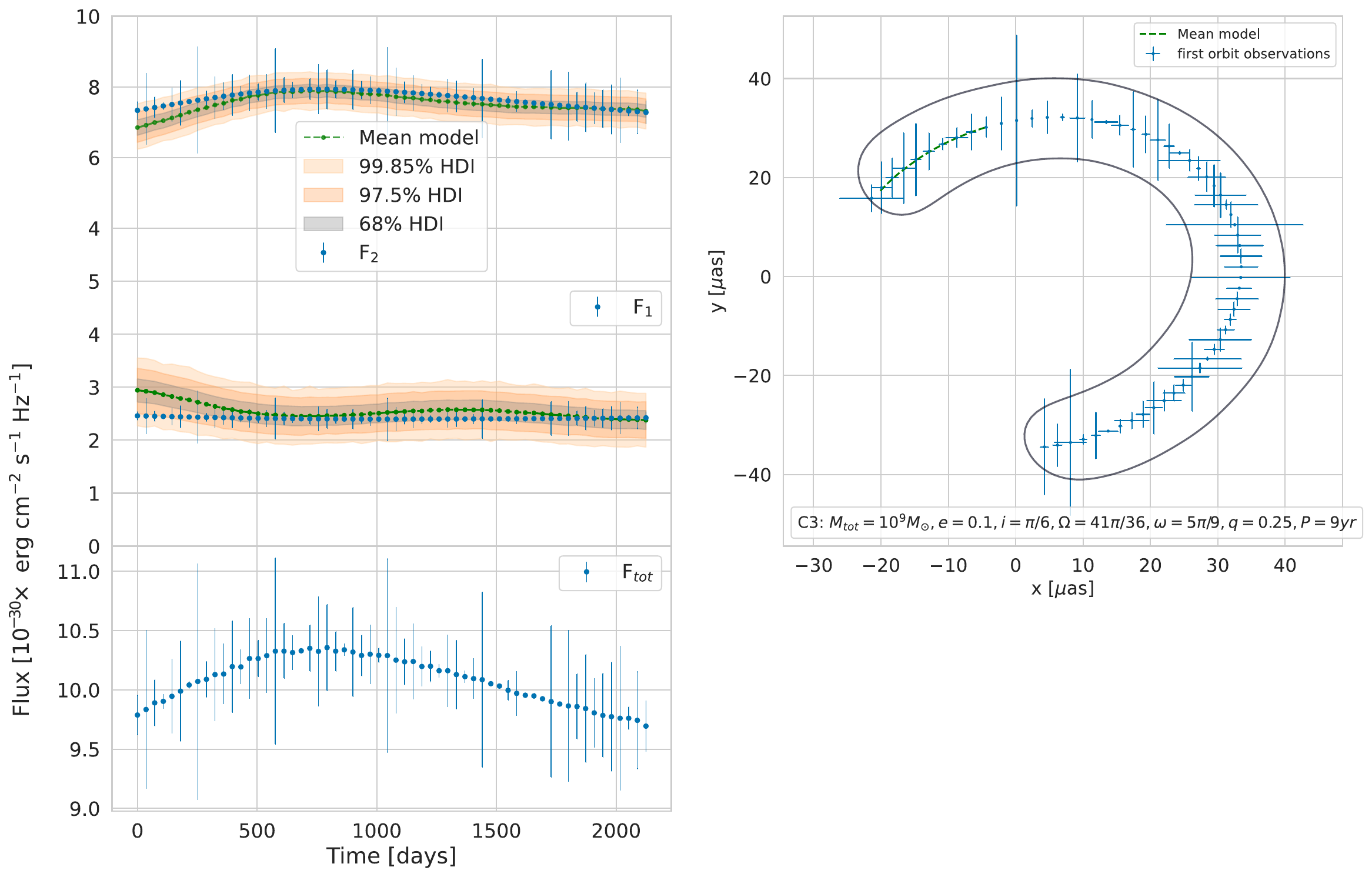}\\
        {(a) $C3$}
    \end{minipage}
    
    % Second image and subcaption
    \begin{minipage}{\textwidth}
        \centering
        \includegraphics[width=\textwidth]{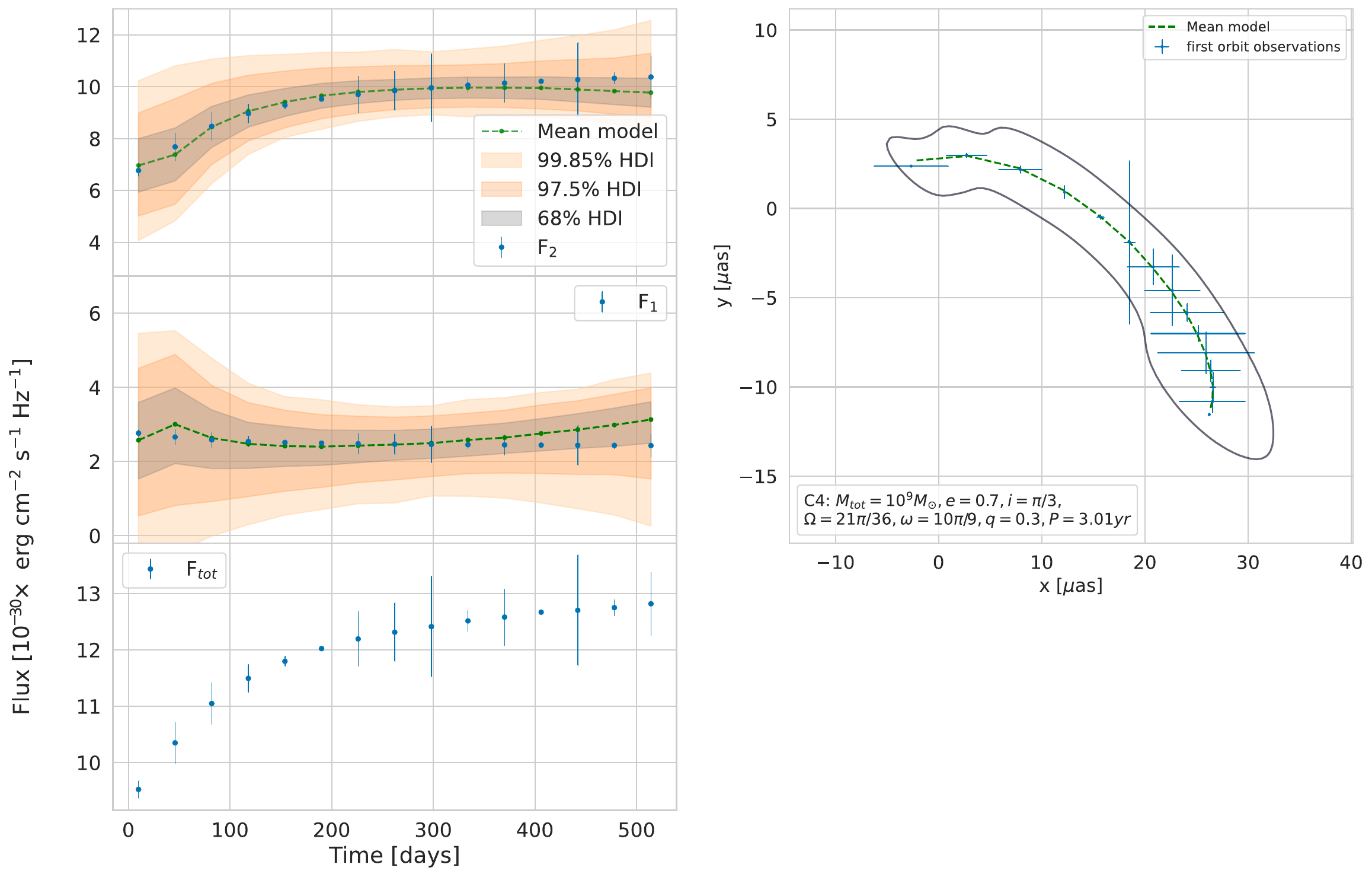}\\
        {(b) $C4$}
    \end{minipage}
    \caption{Bayesian inference of binary system dynamics through disentangled fluxes and astrometric wobble for  configurations (a) $C3$ and (b) $C4$. }
        \label{fig:Fig2}
\end{figure}

\begin{figure}[p]
    \centering
    % First image and subcaption
    \begin{minipage}{0.98\textwidth}
        \centering
        \includegraphics[width=0.98\textwidth]{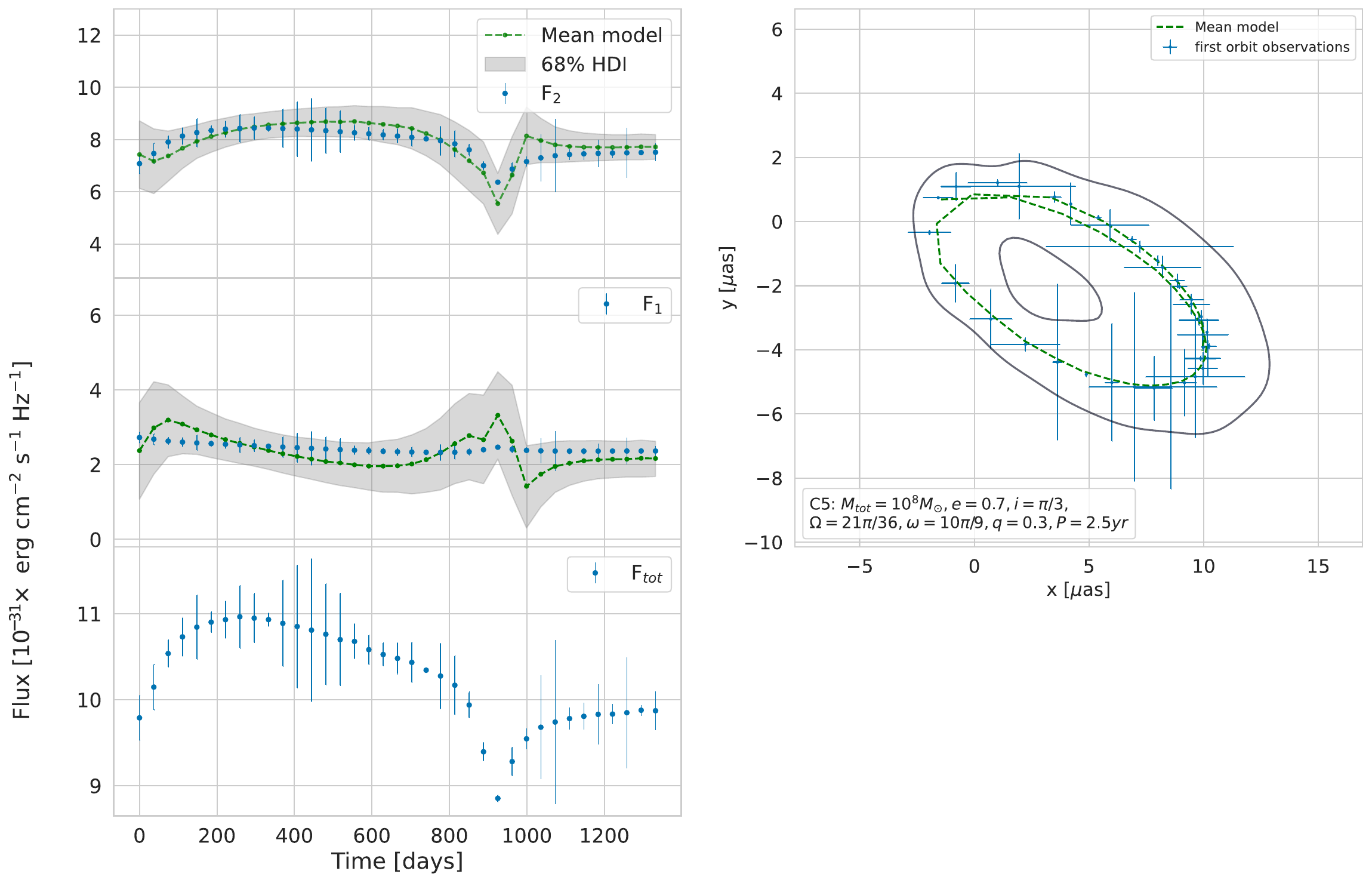}\\
        {(a) $C5$}
    \end{minipage}
    
    % Second image and subcaption
    \begin{minipage}{\textwidth}
        \centering
        \includegraphics[width=\textwidth]{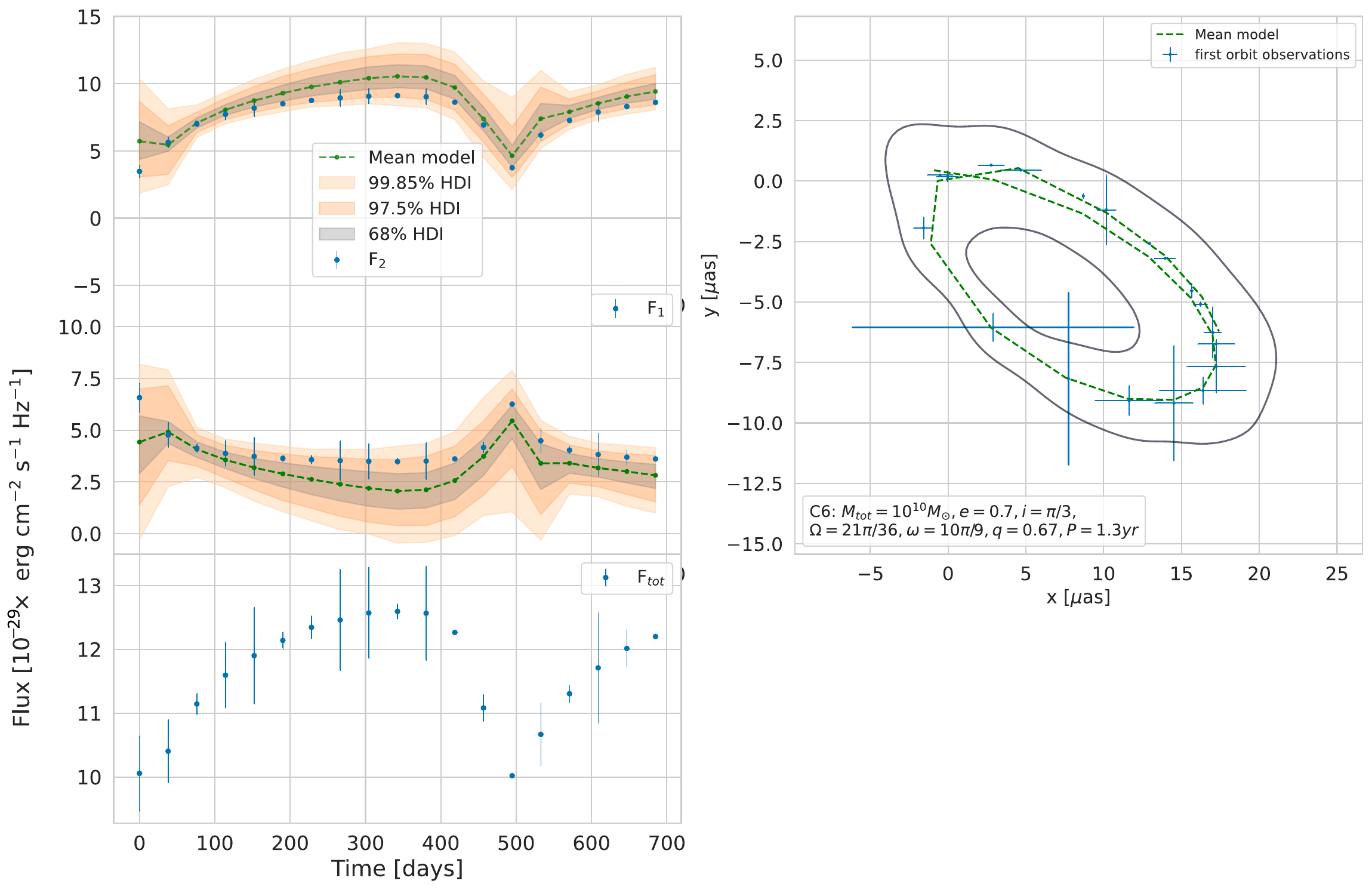}\\
        {(b) $C6$}
    \end{minipage}
    \caption{Bayesian inference of binary system dynamics through disentangled fluxes and astrometric wobble for  configurations (a) $C5$ and (b) $C6$. }
        \label{fig:Fig3}
\end{figure}

To illustrate the procedure (Section \ref{sec:Bayes}) to recover the orbital parameters and dissolve components fluxes,
we consider {the six binary configurations} listed in Table \ref{tab:parameter_combinations} {with varying observational cycles}. 
{Disentangled fluxes of the binary components ($F_{i}, i=1,2$) from the observed total flux $F_{\text{tot}}$ are depicted in Figures \ref{fig:Fig1}-\ref{fig:Fig3}. 
The Bayesian Simulator and Solver employed  the \texttt{PyMC} package, which performs Hamiltonian Monte Carlo (HMC) sampling.
For sampling, in both tiers  of Bayesian code  we  applied the  \texttt{pymc\_ext.optimize} non-linear optimization framework for parameter inference \citep[see details in][]{exoplanet}, which is specifically written for astrophysical applications. It allows us to group astrometric, and emission parameter together, thereby speeding up sampling. We used 4 chains, tuning each for 1000 steps before sampling for a further 40000 steps. The sample have effective sample sizes in the tens of  thousands.  We report 68\% Highest Density Interval (HDI) for posterior values of inferred parameters.
In cases such as C3, C4, and C6,  we include wider $95\% $ and $99.85\% $  HDI to avoid underestimating the uncertainty.
These intervals  align with traditional measures of  1,2 and $3\sigma$, respectively. A simplified notation 1, 2, and 3 HDI denotes the 68\%,  95 \% , and $99.85\%$ HDI, respectively\footnote{The translation from 1 to 3 HDI coverage to conventional sigma ($1-3 \sigma$) levels is based on the equivalence of probability mass within the HDI in Bayesian statistics to the cumulative probabilities associated with sigma levels in a Gaussian distribution. Specifically, a 68\% HDI is analogous to 1$\sigma$, representing approximately 68.27\% coverage, a 95\% HDI corresponds to 2$\sigma$ ($\sim 95.5\% $ coverage), and a 99.7\% HDI to 3$\sigma$ ($\sim$ 99.73\% coverage).  We calculated the effective $1-3 \sigma$  using the full width of  HDI.}. 
The convergence of the Bayesian sampler across scenarios C1-C6 was assessed using two key metrics: the Gelman-Rubin statistic (R-hat) and the effective sample size (ESS). In each scenario, R-hat values were consistently below 1.01 for all parameters, indicating strong convergence and suggesting that the chains mixed well, reaching a stationary distribution. Furthermore, the ESS was several thousands for all parameters, demonstrating that the samples were efficient and informative, providing a solid basis for statistical inferences
Figures \ref{fig:Fig11}-\ref{fig:Fig22} display the Bayesian posterior distributions, and Table \ref{tab:bayes_results} details the inferred binary orbital parameters, providing quantitative estimates and their uncertainties. Table \ref{tab:sigma_metrics}  evaluates the estimation accuracy of these parameters by indicating their deviation from true values within 1 to 3 HDI distances.
Configuration $C1$ (Fig. \ref{fig:Fig1}), characterized by $L_{\text{oc}}>1$, illustrates long-term dynamic trends. Configurations $C3$ (Fig. \ref{fig:Fig2}) and $C6$ (Fig. \ref{fig:Fig3}), also with $L_{\text{oc}}>1$, resemble $C1$ but emphasize different aspects: $C3$ explores the impact of varying orbital parameters, while $C6$ represents a "q-accrete" scenario \citep{10.1093/mnras/stz150}. $C2$ (Fig. \ref{fig:Fig1}) and $C5$ (Fig. \ref{fig:Fig2}) are focused on short-term dynamics ($L_{\text{oc}}<1$). $C4$ (Fig. \ref{fig:Fig1}), targeting intermediate timescales with $L_{\text{oc}}\xrightarrow{}1$, offers a balanced perspective between short-term observational baselines and long-term evolution.
In scenario C1, our model  captures the fluxes of binary components, reflecting long-term dynamic trends within a system characterized by moderate eccentricity and a lower mass ratio over a period of 2.28 years. This scenario includes two full observational cycles relative to the orbital period. Scenario C2, characterized by higher eccentricity and a slightly higher mass ratio over a 1-year period with one observational cycle, demonstrates the method's precision in detecting rapid orbital dynamics and flux variations of components within a single cycle, highlighting its ability to capture short-term dynamical behaviors.
Scenario C3, with low eccentricity over an extended 9-year period and $L_{oc}=2/3$
shows the method's effectiveness in analyzing systems exhibiting slow orbital dynamics with limited observational cycles. In scenario C4, our approach accurately captures complex orbital dynamics in a binary system with high eccentricity and a mid-range mass ratio over a 3.01-year period, observed within moderate cycles of $L_{oc}=1/2$
Scenario C5 underlines the method's precision across different mass scales in a system with high eccentricity and a lower total mass over a 2.5-year period, with $L_{oc}=1/2$
 showcasing its ability to decompose fluxes accurately even in sparsely observed cycles. Finally, C6 focuses on a massive binary system with a short period of 1.3 years and $L_{oc}=1.5$ illustrating the method's adaptability in monitoring flux variations of components across more frequent observing cycles in scenarios involving very high binary mass and rapid orbital periods.}

\newpage

\newpage

\begin{table*}[ht]
\centering
\caption{Bayesian inference of binary parameters with 68 \% HDI for  configurations  given in Table \ref{tab:parameter_combinations}.}
    \label{tab:bayes_results}
\begin{tabular}{l|c|c|c|c|c|c|c}
\hline
Configuration& $q$ & $P$ [yr] & $i [^{\circ}] $& $\Omega [^\circ]$ & $\omega [^{\circ}]$ & $M_{\mathrm{tot}} [M_\odot]$ & $e$ \\ 
\hline
C1 & $0.23_{-0.02}^{+0.02}$ & $2.28_{-0.01}^{+0.01}$ & $61.13_{-1.19}^{+1.29}$ & $44.88_{-1.63}^{+1.54}$ & $59.97_{-1.85}^{+1.86}$ & $0.99_{-0.02}^{+0.02} \times 10^9$& $0.30_{-0.01}^{+0.01}$ \\ 
C2 & $0.43_{-0.06}^{+0.05}$ & $1.00_{-0.01}^{+0.01}$ & $28.17_{-11.53}^{+17.22}$ & $46.86_{-33.02}^{+24.76}$ & $58.00_{-18.26}^{+37.34}$ & $1.00_{-0.02}^{+0.02} \times 10^9$ & $0.50_{-0.04}^{+0.04}$ \\ 
C3 & $0.26_{-0.02}^{+0.01}$ & $9.28_{-0.05}^{+0.06}$ & $24.16_{-2.08}^{+2.21}$ & $214.09_{-6.70}^{+4.39}$ & $96.03_{-3.74}^{+6.11}$ & $0.99_{-0.02}^{+0.02} \times 10^9 $ & $0.06_{-0.01}^{+0.01}$ \\ 
C4 & $0.33_{-0.04}^{+0.04}$ & $3.93_{-0.51}^{+0.42}$ & $55.70_{-5.13}^{+8.89}$ & $110.88_{-13.33}^{+6.16}$ & $196.94_{-19.46}^{+27.44}$ & $1.00_{-0.02}^{+0.03}  \times 10^9$ & $0.75_{-0.05}^{+0.08}$ \\ 
C5 & $0.28_{-0.03}^{+0.03}$ & $2.60_{-0.02}^{+0.02}$ & $64.81_{-2.67}^{+3.92}$ & $99.12_{-3.04}^{+4.21}$ & $216.12_{-6.08}^{+5.62}$ & $1.00_{-0.02}^{+0.02}  \times 10^8$ & $0.72_{-0.04}^{+0.03}$ \\
C6 & $0.69_{-0.04}^{+0.05}$ & $1.39_{-0.02}^{+0.03}$ & $69.49_{-3.10}^{+4.20}$ & $95.38_{-8.34}^{+6.75}$ & $226.01_{-6.45}^{+16.21}$ & $0.95_{-0.13}^{+0.09}  \times 10^{10}$ & $0.76_{-0.07}^{+0.10}$ \\ \hline

\hline
\end{tabular}
\end{table*}

\begin{table}[ht]
\centering
\caption{Symbolic summary of Bayesian inference results  for each configuration given in Table \ref{tab:parameter_combinations}, indicating whether the estimated orbital parameters are within 1-3 HDI distance from the true values.}
\label{tab:sigma_metrics}
\begin{tabular}{|c|c|c|c|c|c|c|c|}
\hline
Configuration & $e$ & $q$ & $i$  & $\Omega$  & $\omega$  & $P$  & $M_{\text{tot}}$  \\
\hline
C1 & 1 HDI & 1 HDI & 1 HDI & 1 HDI & 1 HDI & 1 HDI & 1 HDI \\
C2 & 1 HDI  & 1 HDI & 1 HDI & 1 HDI & 1 HDI & 1 HDI & 1 HDI \\
C3 & 1 HDI & 1 HDI & 1 HDI & 1 HDI & 1 HDI & 2 HDI & 1 HDI \\
C4 & 1 HDI & 1 HDI & 1 HDI & 1 HDI & 1 HDI & 2 HDI & 1 HDI \\
C5 & 1 HDI & 1 HDI & 1 HDI & 1 HDI & 1 HDI & 1 HDI & 1 HDI \\
C6 & 3 HDI & 2 HDI & 1 HDI & 1 HDI & 1 HDI & 2 HDI & 2 HDI \\
\hline
\end{tabular}
\end{table}

\section{Discussion}\label{sec:discussion}
Here, we discuss observational strategies,  the expected astrometric wobble distributions within the parameter space of CB-SMBH and address potential limitations in our methodology.

\subsection{Comparison of Observational Strategies}

{
Our Bayesian synthesis method successfully estimated the orbital parameters of the binary quasar system across six configurations (Table \ref{tab:parameter_combinations}), demonstrating the model's capability to accurately recover parameters (Table \ref{tab:bayes_results}). The estimated values are generally in close agreement with the true values (Figure \ref{fig:Fig11}-\ref{fig:Fig33}, with most parameters falling within the  1HDI  and confidence intervals ranging from 1HDI to, in a few cases, 2-3HDI for more challenging configuration of "q-accrete" C6 (Table \ref{tab:sigma_metrics}).
Configuration C1 benefits from an optimal match between observing cycles and the orbital period, supported by moderate eccentricity and mass ratio. This balance enables precise components fluxes decomposition within $68\%$ HDI, highlighting the effectiveness of the observational strategy in capturing the binary's stable dynamics over two complete cycles.
C2 demonstrates the model's ability to accurately capture rapid dynamical changes within a shorter, 1-year period with a single observing cycle. The higher eccentricity here necessitates dense observational coverage, which is achieved, resulting in flux decompositions aligning within $68\%$ HDI.
C3 faces challenges due to its extended 9-year period and a lower $L_{\text{oc}}=2/3$, which complicates the full capture of long-term dynamics, placing decomposed fluxes within $97.5\%$ at the beginning of observing cycle. This reflects the difficulty in tracking the binary's evolution with sparser data points.
C4 and C5 contend with high eccentricity ($e\sim 0.7)$, impacting the flux decomposition's precision. The longer period and half-cycle observing rates introduce complexities in fully capturing the binary's dynamics, likely resulting in decomposed fluxes aligning within $97.5\%$ sigma in the case of C4. However, due to smaller period of C5 the cadence is enough granular to decompose fluxes within $68\%$ HDI.Additionally, the relatively higher eccentricity and lower total mass in C5 compared to other configurations may contribute to more distinguishable flux variations that are easier to capture within the given observational sampling.
C6, with a higher $L_{\text{oc}}$ and a significant mass  of 'q-accrete' within a compact 1.3-year period, faces intense dynamical variations. The alignment of decomposed fluxes within wider sigma bands for C6 indicates the model's capacity to account for these rapid variations to a certain extent. However, the deviations from tighter sigma bands underscore the necessity for more nuanced observational strategies, such as increasing the sampling rate during critical phases or employing predictive modeling to enhance the temporal resolution of observations.
Given the uncertainty surrounding the specific type of binary system at the outset of observation, the  configurations examined here point out the importance  of a flexible and adaptive observational approach, capable of accommodating the wide range of binary characteristics that might be encountered:
\begin{itemize}
    \item 
Start with a baseline sampling rate, such as the 10 observations per year, and adjust based on early data analysis. If rapid flux variations or complex orbital dynamics are detected, increase the observation frequency accordingly. 
\item  Multi-wavelength observations can provide additional clues about the binary system's nature, aiding in the refinement of observational strategies to better target the detected features. 
\item Collaborations that allow for shared data and observations from different facilities and instruments can enhance the detection capabilities and flux decomposition.
\end{itemize}
An example of workflow could involve collecting LSST catalogue data on variability,  and providing this information to future astrometry facilities. These facilities can then use the catalogue to optimize their observing strategy in real time based on modeling, ensuring efficient capture of dynamic binary behavior.}

\subsection{Distributions of Expected Astrometric Wobble}

Here we predict and examine the astrometric wobbles of CB-SMBH across two distinct parameter spaces: the $(q/(1+q), l)$ plane, and the total mass-period domain. Within each space, we incorporate variations in eccentricity and orbital phase. The purpose of the predictions is to assist and enhance multi-instrument observational strategies.

For the computations of the offset across a grid of binary parameters, a simplified  equation Eq. \ref{eq:photonewx} is employed, wherein the luminosity parameter $l$ is assumed to be time-invariant, and both the luminosity and mass ratios are presumed to have a uniform distribution. This assumption allows us to focus exclusively on the dynamical parameters of the orbit—namely, the semi-major axis $a$, eccentricity $e$, true anomaly  $\nu$\footnote{Equivalent term is orbital phase.}, inclination $i$, and orbital orientation $\omega$. The formula for total photocenter displacement could have positive or negative sign  depending on the values of $\mu$ and $l$. If $\mu>l$, then the component with the larger mass contributes less to the total light of the system, and vice versa. 

Given that the most significant effects are expected from objects within tens of Mpc \citep{Dexter_2020}, we set the distance of the object to be roughly 70 Mpc. We then compute the offsets (Eq. \ref{eq:photonewx}) for various binary parameters combinations in the $(q/(1+q), l)$ plane, under the assumption of a fixed inclination $i=0$.

\begin{sidewaysfigure}
\centering
\begin{tabular}{ccc}
\includegraphics[width=0.33\textwidth]{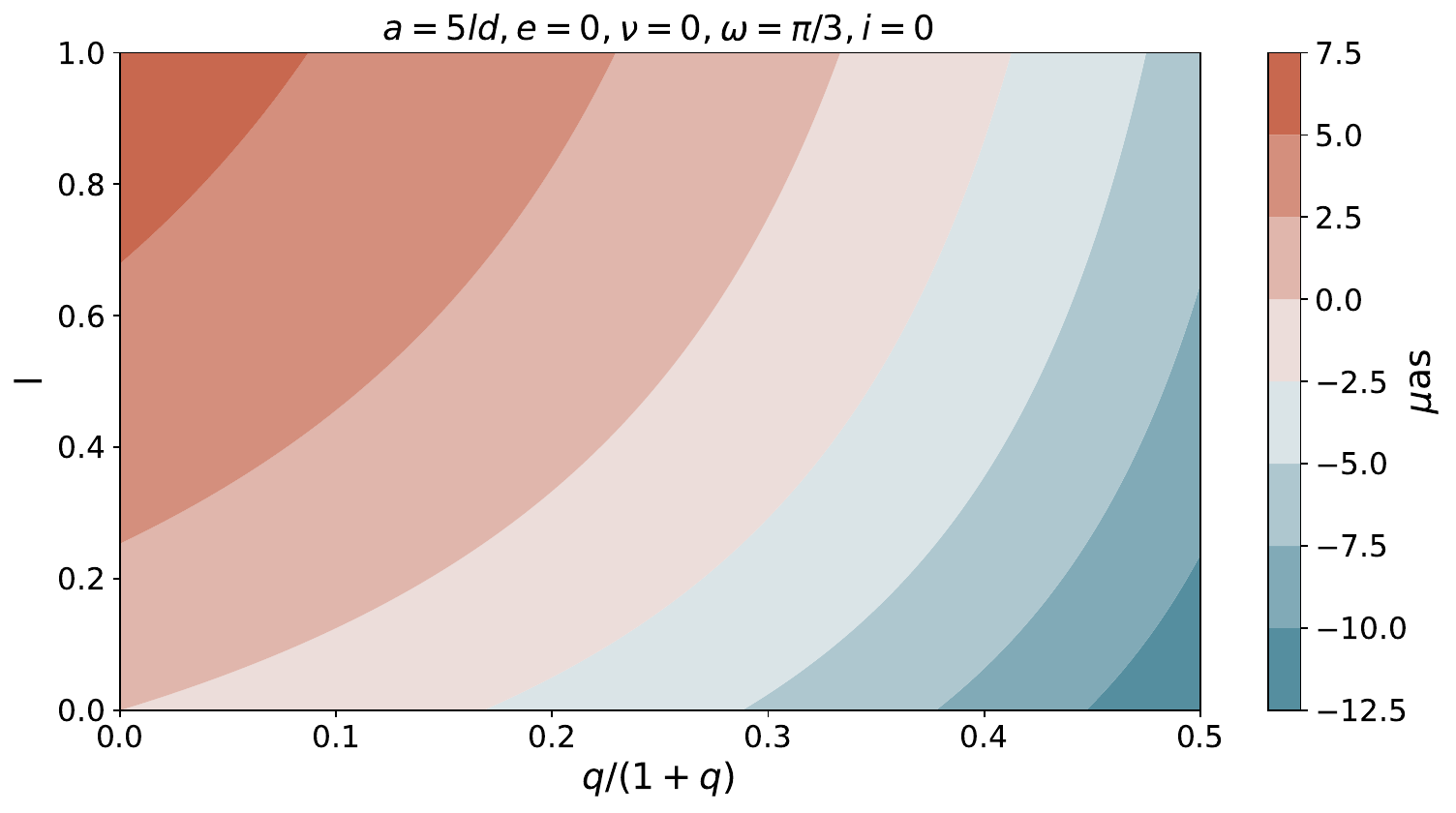} & 
\includegraphics[width=0.33\textwidth]{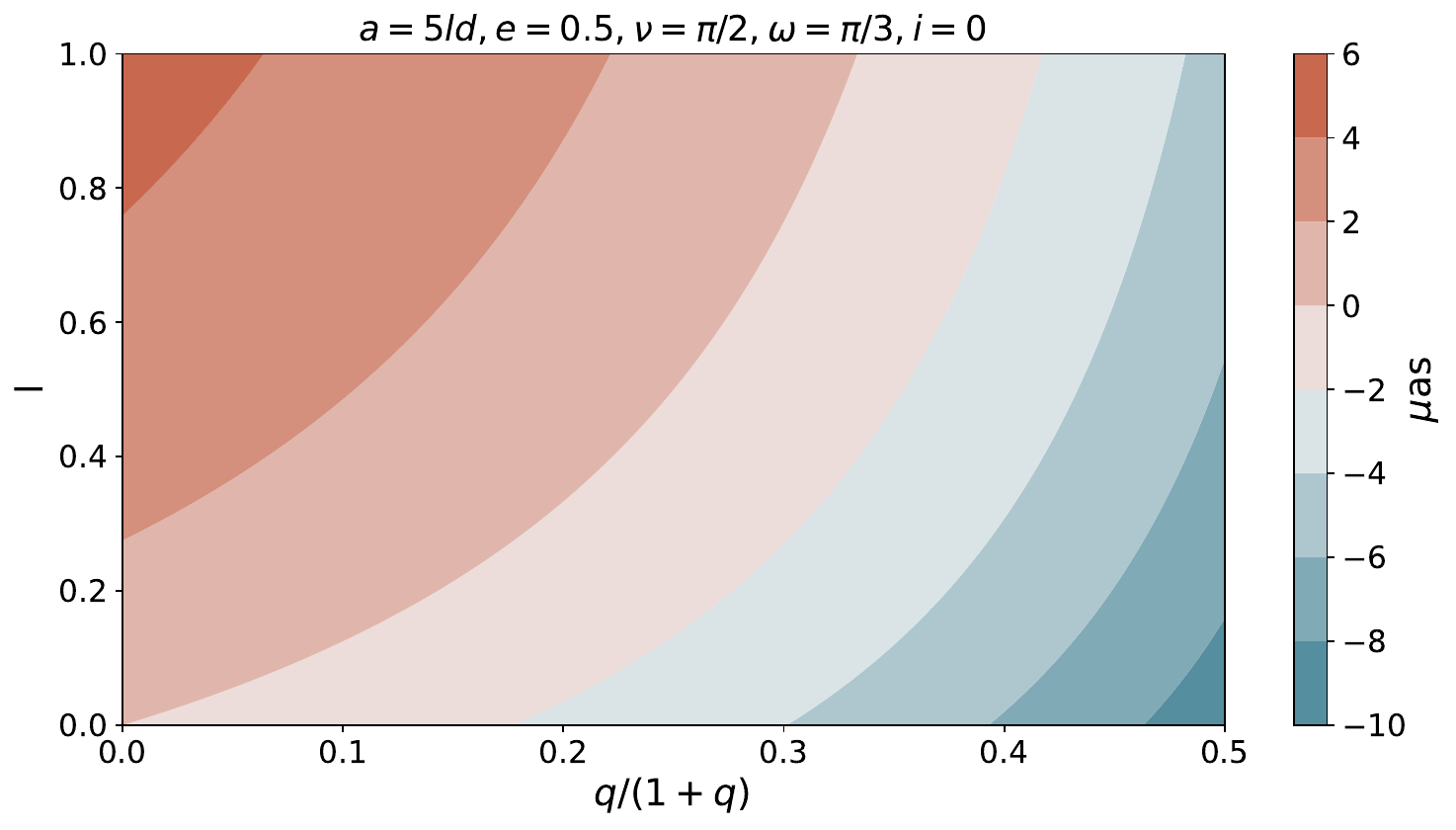} & 
\includegraphics[width=0.33\textwidth]{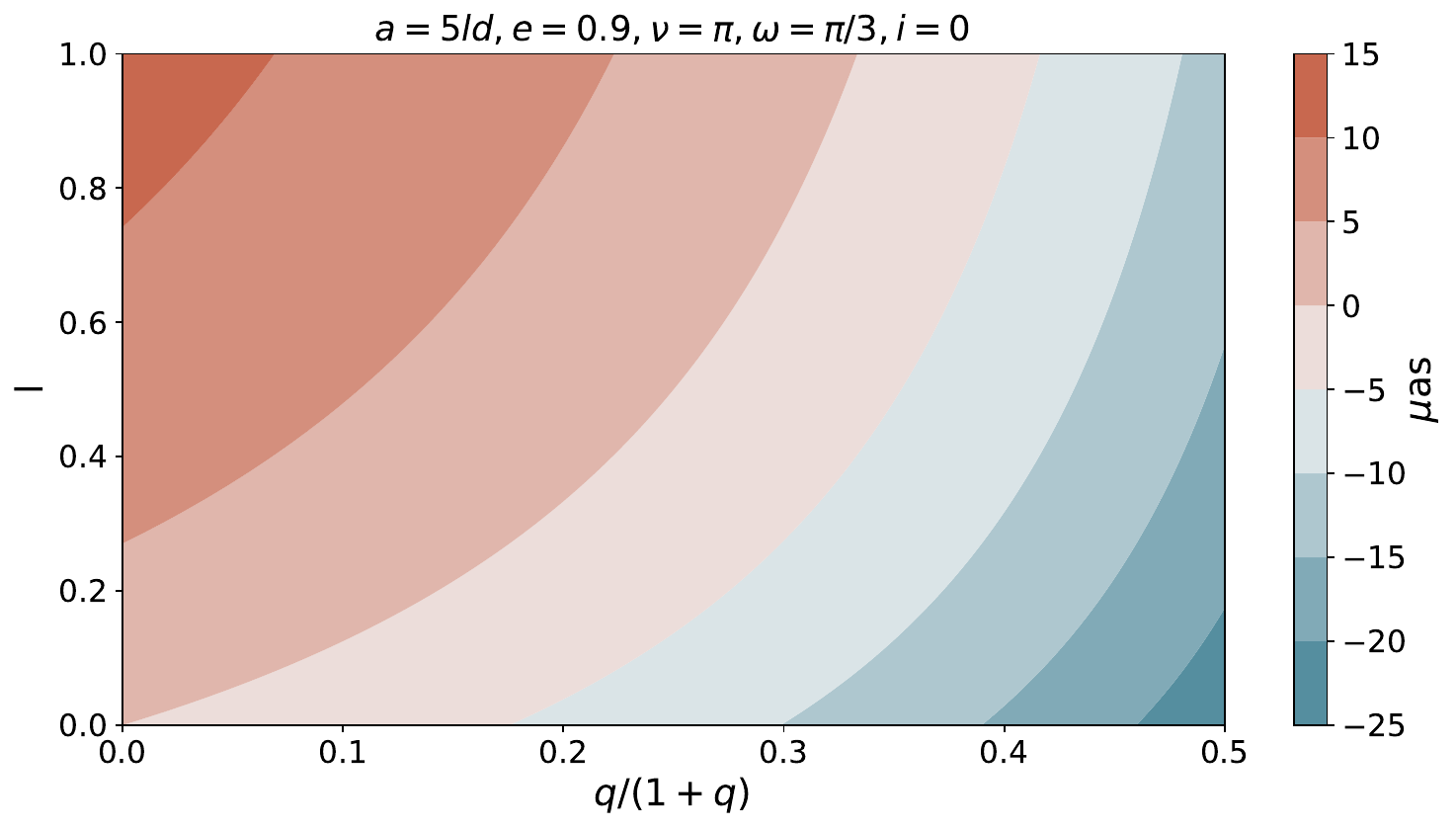} \\
(a) & (b) & (c) \\
\includegraphics[width=0.33\textwidth]{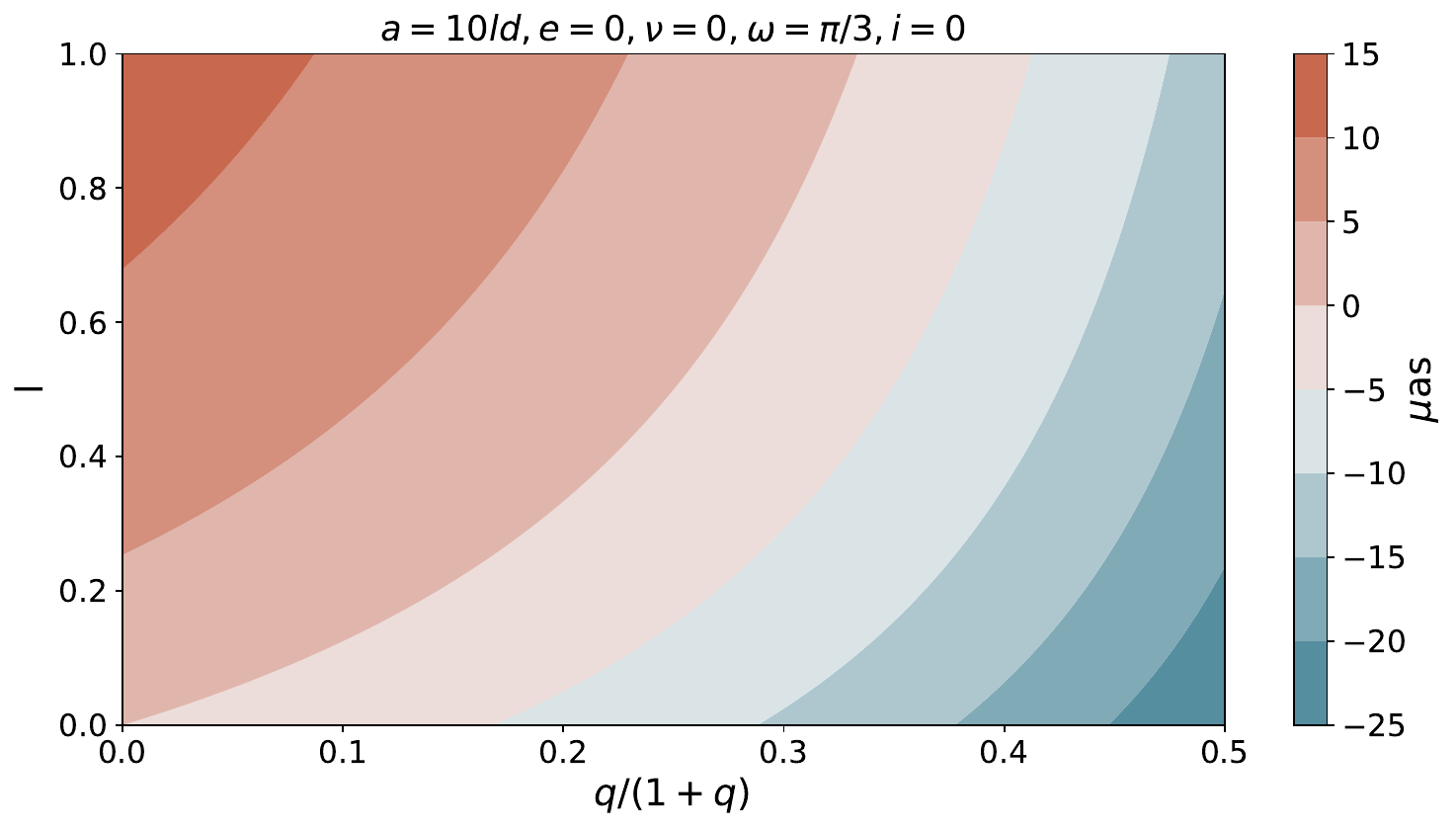} & 
\includegraphics[width=0.33\textwidth]{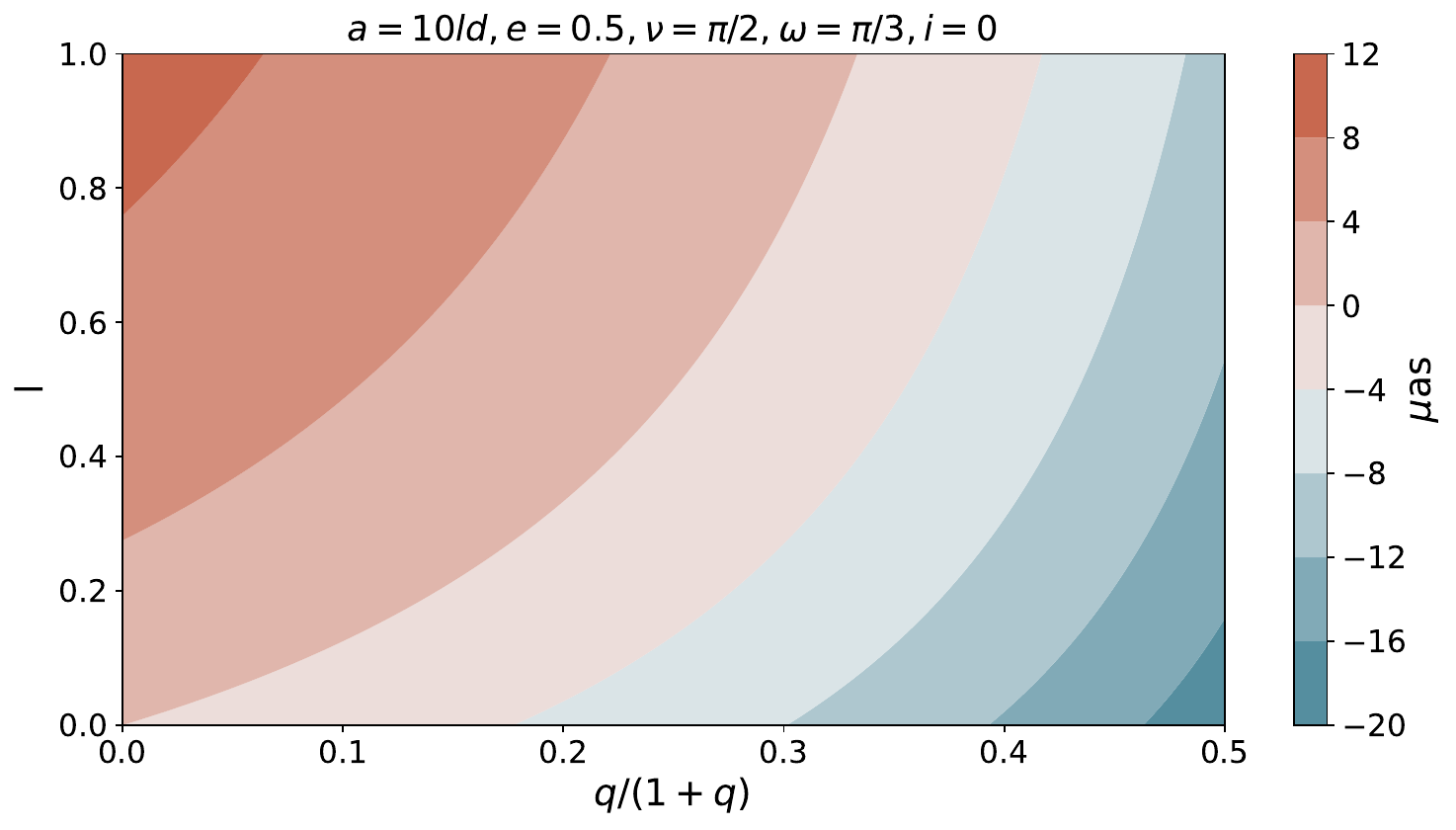} & 
\includegraphics[width=0.33\textwidth]{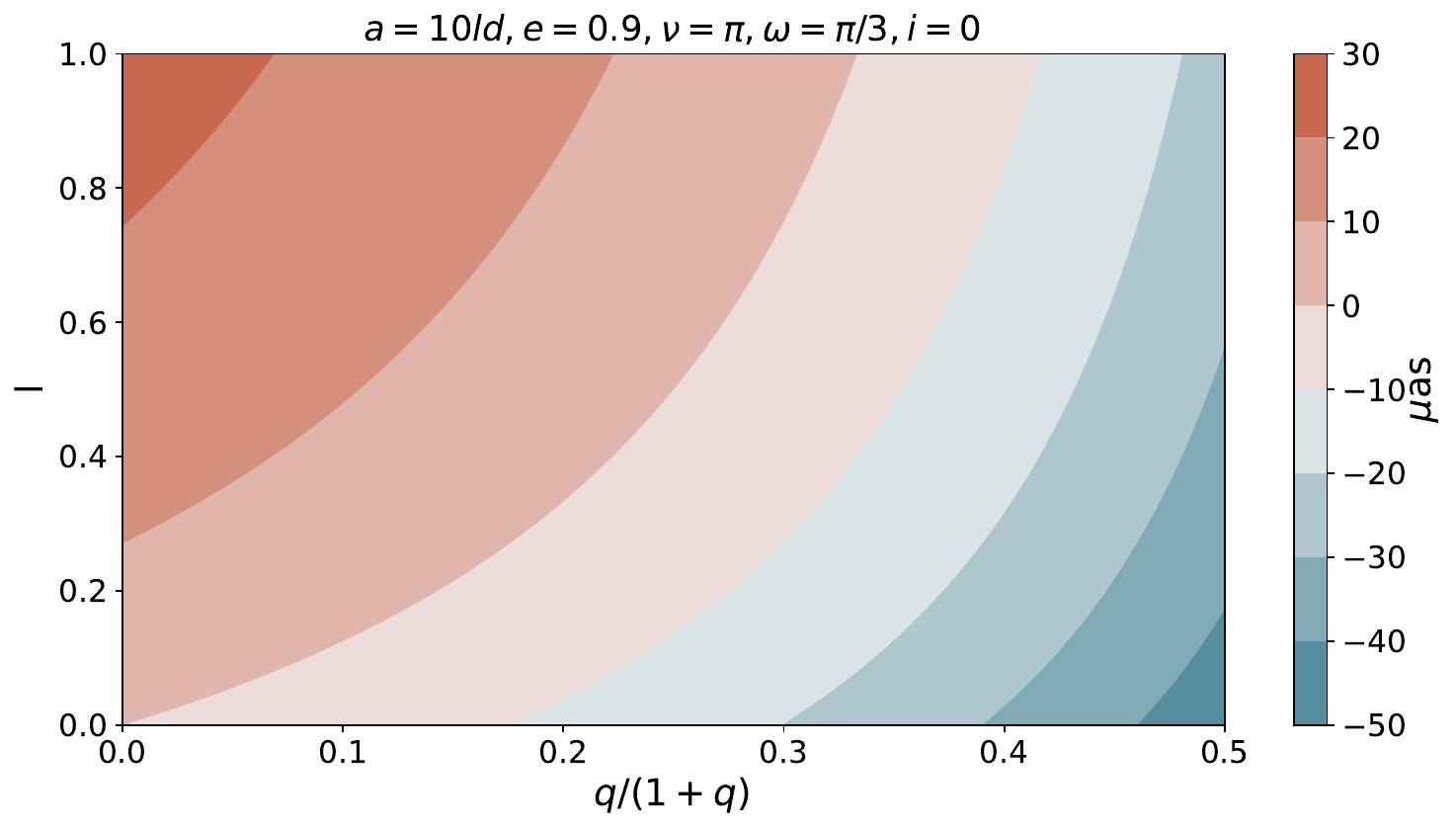} \\
(d) & (e) & (f) \\
\end{tabular}
\caption{Predicted astrometric wobble based on a grid of parameters, for varied $q/(1+q)$ and $l$ values. Color gradient depicts wobble  magnitude and direction. Both rows of plots maintain fixed inclination $i=0$, and  argument of pericenter $ \omega=\pi/3$. The top row, set at semi-major axis $a=5$ ld, presents: $e=0$, $\nu=0$ (a), 
 $e=0.5$, $\nu=\pi/2$ (b),
 $e=0.9$, $\nu=\pi$ (c). The bottom row, set at $a=10$ ld, shows:  $e=0$, $\nu=0$ (d),  $e=0.5$, $\nu=\pi/2$ (e),
 $e=0.9$, $\nu=\pi$ (f).}
\label{fig:general}
\end{sidewaysfigure}

In Figure \ref{fig:general}, the distinction between offsets for different values of the semi-major axis 
$a$ is clear. In the top row, where 
$a=5$ ld, the offsets are consistently lower than in the bottom row, with 
$a=10$ ld, emphasizing the influence of semimajor axis on the offsets.

The variation in eccentricity and true anomaly across the columns of Figure \ref{fig:general} illustrates the influence of these parameters on the resultant offsets. The uniformity in the first column (subplots a, d) for a circular orbit contrasts with the pronounced variation observed in the second (subplots b, e) and third columns (subplots c, f), especially at higher eccentricities.  At a moderate eccentricity 
$e=0.5$  and a true anomaly 
$\nu= \pi/2$, the offset variation is more pronounced, especially when comparing it with the circular case in the first column. The slight differences between subplots (b) and (e) further highlight the enhanced offset sensitivity to the orbital phase at this eccentricity.
With a high eccentricity $e=0.9$ and the orbital phase close to  $\pi$, the offset values display a  contrast, in both subplots (c) and (f) with respect to other cases. This  underscores the pronounced effects of having components near the apocenter of their elliptical orbit, producing more substantial displacements.

\clearpage 

\clearpage

\begin{figure}
\centering
\includegraphics[width=0.5\textwidth]{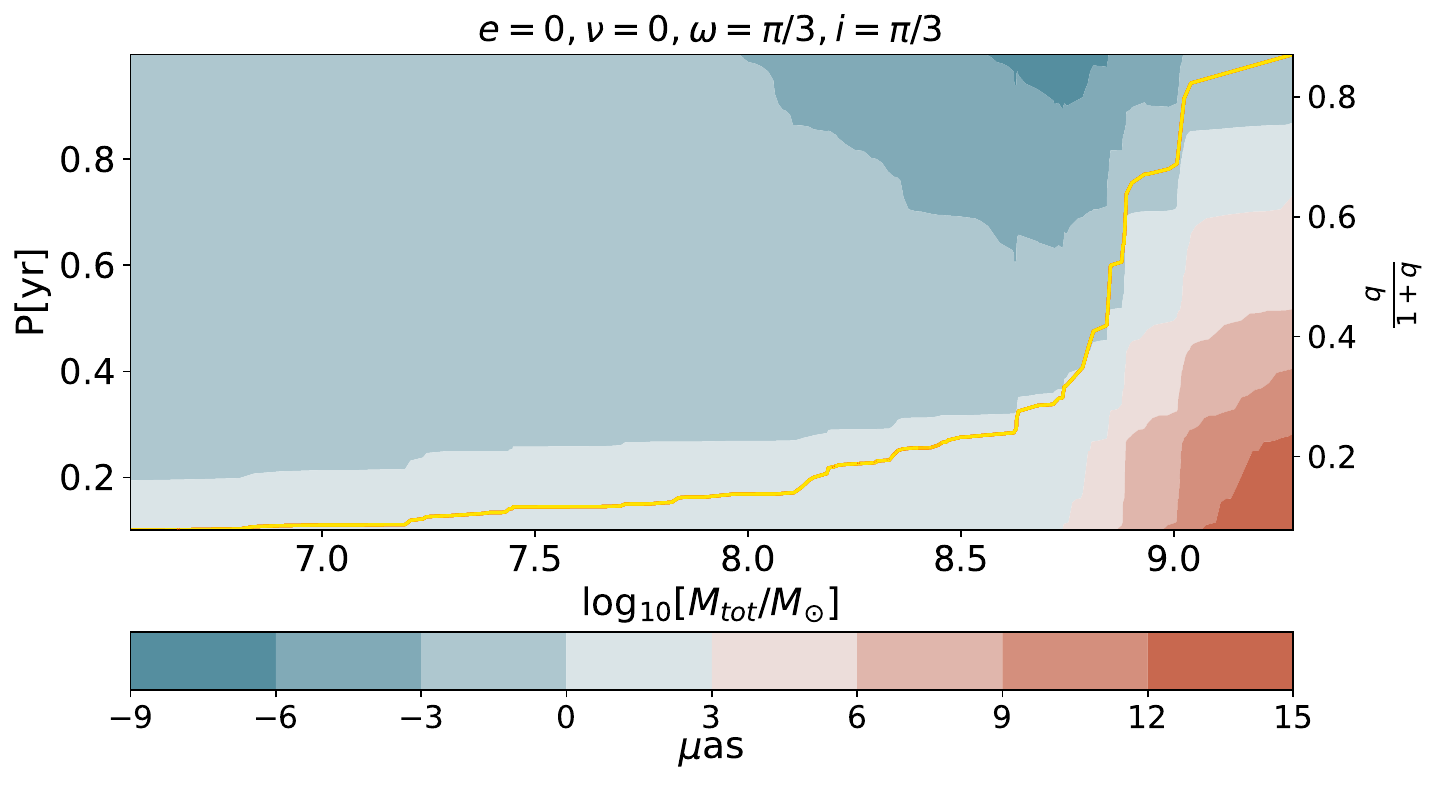} \\
\makebox[0.5\textwidth][c]{(a)}\\[-0.5ex]

\includegraphics[width=0.5\textwidth]{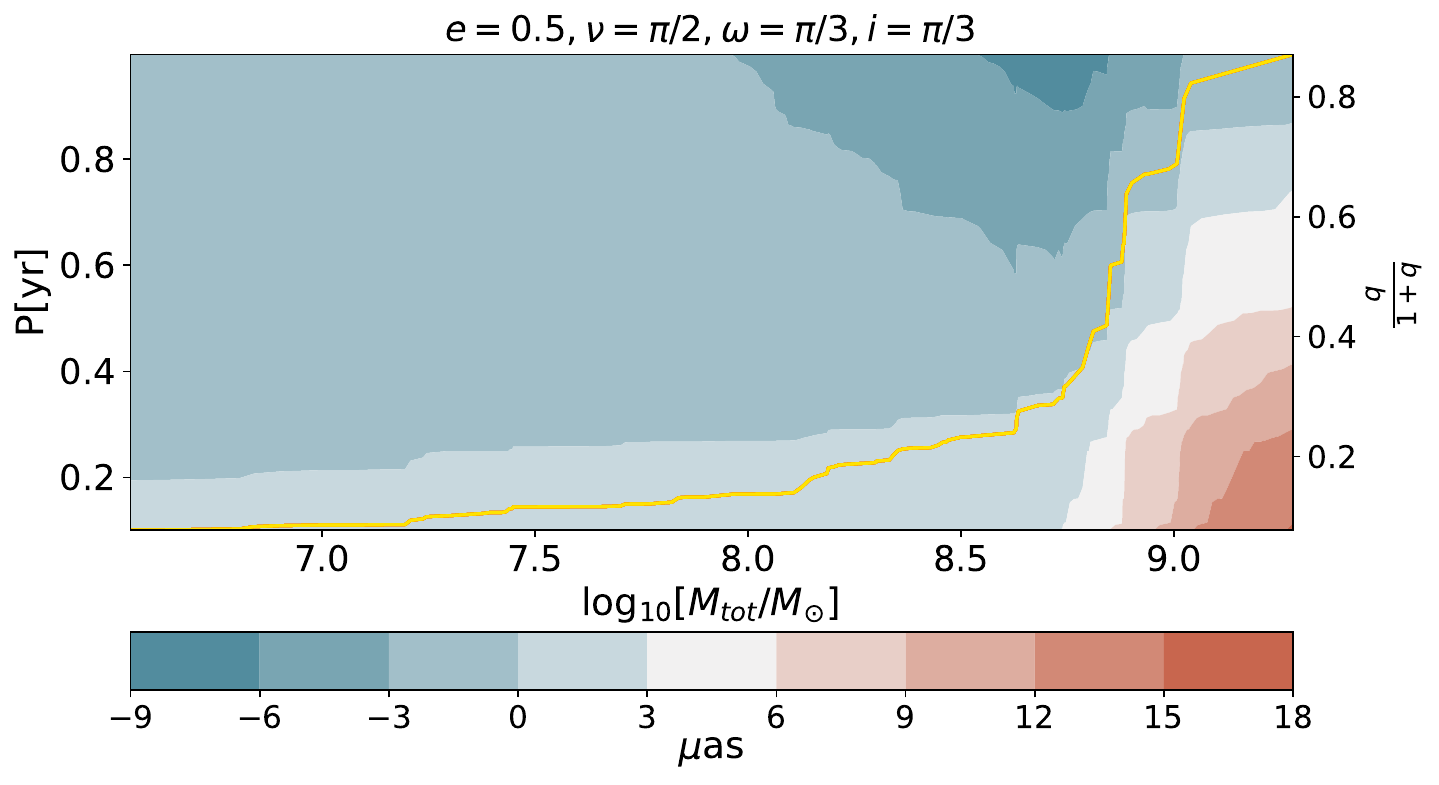} \\
\makebox[0.5\textwidth][c]{(b)}\\[-0.5ex]

\includegraphics[width=0.5\textwidth]{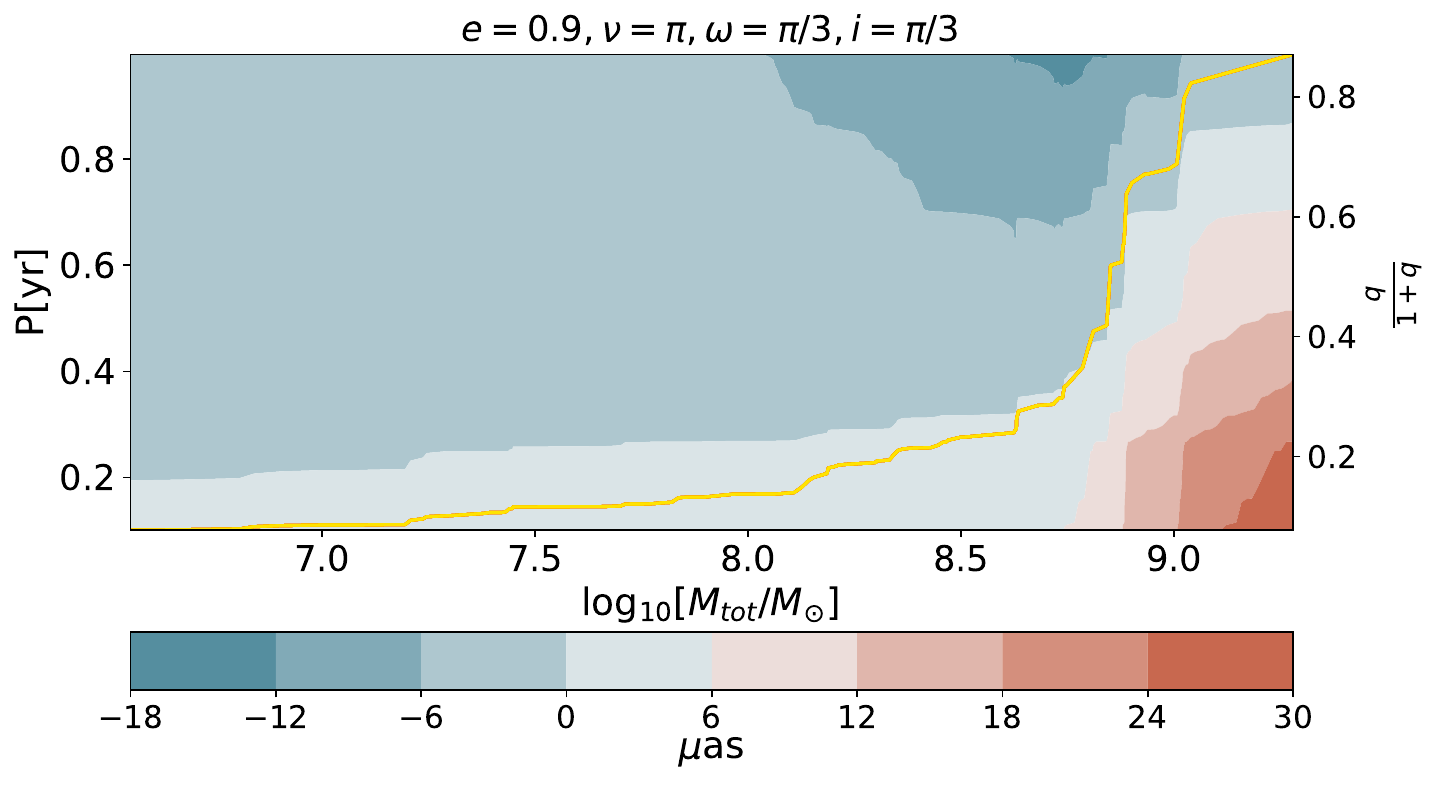} \\
\makebox[0.5\textwidth][c]{(c)} %\\[-4ex]
\caption{ Predicted astrometric wobble based on varying total mass and period of CB-SMBH. All plots have consistent $\omega=\pi/3, i=0$, but differ in: $e=0,\nu=0$ (a); $e=0.5,\nu=\pi/2$ (b); $e=0.9,\nu=\pi$ (c). Color gradient and yellow curves indicate offset magnitude and direction, and barycenter parameter $q/(1+q)$, respectively.  }
\label{fig:general2}
\end{figure}

We now focus on the photocenter displacement,$\lbrack \bm{\delta}{ph}\rbrack$, as a function of the total mass and period, $\lbrack \bm{\delta}{ph}( \log[M_{tot}/M_{\sun}], P)\rbrack$ (see Figure \ref{fig:general2}). This alternative phase plane is crucial due to its direct link to the binary's physical and  orbital parameters.
For this analysis, we reparametrize the binary's relative position, $r$, in Eq. \ref{eq:photonewx} using the formula:
\begin{equation}
r=\frac{a(1-e^{2})}{1+e\cos\nu}=\left(P^{2}M_{tot}\right)^{1/3}\frac{1-e^{2}}{1+e\cos\nu}
\end{equation}
This equation shows the dependencies on the total mass $M_{\text{tot}}$, orbital period $P$, eccentricity $e$, and the true anomaly $\nu$. In an eccentric orbit, the radius $r$ fluctuates, peaking at $\nu = \pi$ (the apocenter). Hence,  the maximum displacements may occur when the binary system's orbit is highly eccentric ($e=0.9$), and the orbital phase, defined by the true anomaly $\nu$, is at $\pi$. This aligns with the apocenter, where the two bodies in the binary system are maximally separated.

 Based on Figure \ref{fig:general2}, we outline several trends dictated by dynamical parameters. In the  top subplot,  for  circular systems encompassing total masses within  
$10^{7}$ to $10^{8}M_{\sun}$ 
 , there exists a balanced distribution of positive and negative offsets. However, this  gets disrupted for masses in the range from $10^{8}$ to $10^{9}M_{\sun}$, wherein a pronounced tendency to positive offsets emerges.  This suggests that as we progress to more massive systems, the distinction between the light-weighted center and mass-weighted center becomes more pronounced.

 Increasing the eccentricity to 0.5 introduces deviations from circularity in the orbits. This scenario is set at $\nu=\pi/2$. Here, the positive offset regions have enlarged compared to top subplot, especially for higher mass systems. This trend underscores the perturbative effects of higher eccentricities: as orbits deviate from perfect circles, gravitational interactions over the orbit become more varied, leading to pronounced photocenter displacements.
The bottom subplot captures the systems at the most elliptical scenario, with the  diametrically opposed components positions. The significant eccentricity profoundly impacts the offset distributions. The positive offset region has not only enlarged but also stretched towards higher masses. Notably, the boundaries defined by the yellow curve, representative of the barycenter parameter 
 are most prominent here, illustrating how the offset distributions are inherently tied to this parameter, especially at such high eccentricities.

The plane of $(\log[M_{tot}/M_{\sun}], P)$ reveals that the demarcation between positive and negative total offset values are delineated with the mass ratio $\mu=q/(1+q)$. Specifically, if $\mu>l$, then the component with the larger mass contributes less to the total emission of the system, and the total photocenter displacement $[\bm{\delta}{ph}]$ will be positive. This may suggest that the more massive component is less luminous (or active) or possibly obscured. On the other hand, if $\mu<l$, then the component with the larger mass contributes more to the total light of the system, and the photocenter displacement $[\bm{\delta}{ph}]$ will be negative. This may suggest that the more massive component is more luminous (active) or less obscured. This relationship is particularly pronounced for total masses exceeding $10^{8}M_{\sun}$ and periods $\geq 1$ year, as larger masses and extended periods lead to larger orbital separations and, consequently, more significant photocenter displacements. 
Since our method provides the mass ratio  and disentangles the light curve of the secondary, total photocenter enables us to derive a preliminary indicator of the activity of the  components in the CB-SMBH system.

\subsection{Caveats}
 Here we highlight further aspects that can modulate the interpretation of our results.
\begin{itemize}
\item 
{\it Orbital Phase Dependancy}: The photocenter offset exhibits a  dependence on the orbital phase, with smaller variations at the pericenter compared to the apocenter. This could be seen from the ratio of photocenter offset values in pericenter $\Delta_{p}$ and apocenter $\Delta_{ap}$ inferred from Eq.\ref{eq:photonewx}:
\begin{equation}
    \frac{\Delta_{p}}{\Delta_{ap}}=\frac{1-e}{1+e}
\end{equation}
However, we also see  that for CB-SMBH systems with mild eccentric orbits ($0 < e < 0.3$), the offset ratio tends to approach unity, implying that even at pericenter, the observed offset can be relatively nonneglibile. This phase-dependent behavior necessitates careful consideration when interpreting observed offsets, especially for systems with varying eccentricities \citep[see related discussions in][]{2022A&A...663A..99K}.
\item {\it Parallax and Proper Motion}: Our analysis did not account for the possibility of parallax and proper motion. While we assumed that most short-period CB-SMBH systems exhibit small and likely imperceptible parallax, still,  we cannot exclude the existence of systems with larger parallax. Furthermore, proper motion of quasars may arise from various sources, such as differential variability between close sources \citep[see][]{Makarov_2022}, or due to the motion of a relativistic jet, as in the case of PKS 0119+11 \citet{Lambert21}, which could affect proper motion measurements. These factors merit further investigation and consideration when interpreting photocenter offset measurements \citep[see][]{Makarov_2022, Lambert21}.
According to \citet{Makarov_2022} many proper motion quasars may be unresolved dual systems displaying the variable imposed motion effect, while a smaller number may be coincidence alignments with foreground stars creating weak gravitational lensing.
\item {\it Low-Mass Galaxies with Black Holes}: Our study primarily focused on massive CB-SMBH systems; however, it is essential to recognize that low-mass galaxies hosting black holes with masses approaching $\sim 10^{5}M_{\sun}$ may require a separate examination. Several campaigns in recent years \citep[see e.g.,][]{Reines2013, Molina2021} have shown that 50–80 percent of low-mass galaxies $M_{g}\sim 10^{9}-10^{10} M_{\sun}$ may harbor in their core low mass black holes with masses approaching  $\sim 10^{5}M_{\sun}$ \citep[see][and references therein]{10.1093/mnras/stab1856}.
These black holes follow scaling relations similar to SMBHs \citep[see e.g.][]{10.1111/j.1365-2966.2009.15577.x} and could impact the study of binary candidates, given their unique characteristics and potential contribution to active galactic nuclei \citep[see][]{Kimbrell_2021}.
\item {\it Large Binary Separations and Disc Alignment}: The dynamics of CB-SMBH systems may change significantly at large binary separations or periods. Circumbinary gas accretion can dominate over gravitational wave emission, and the alignment of circumbinary discs with the orbital plane introduces complexities that deserve further investigation. Such systems may require a distinct treatment within the CB-SMBH framework \citep[see][]{2015MNRAS.449...65A, 10.1093/mnras/stx2936, 2013ApJ...774...43M}.
For example, any periodic emission from the binary mini-discs or the circumbinary disc's inner edge will be obscured if the binary orbit is co-planar with a circumbinary disc.
\item {\it Variability of Accretion Discs}: In cases where the source periodicity is primarily driven by the variability of the accretion disc rather than the Doppler boost, modeling the motion of the photocenter becomes challenging. The resulting curve may exhibit substantial noise and not form a closed ellipse, affecting our ability to model and interpret photocenter motion accurately \citep[see also discussion in][]{2022A&A...663A..99K}.
\item {\it Detection of Ultra-Compact Binaries}: Extremely ultra-compact binary systems with very short periods \citep[see details in][]{10.1093/mnras/stab1856} may present challenges for astrometric detection. The expected offsets for such systems may fall below the detection threshold for current astrometric instruments \citep[e.g., GAIA threshold is $10\mu$as, see][]{PhysRevD.100.103016}. Therefore,  time-domain optical surveys like LSST and future gravitational wave observatories like LISA \citep{2017arXiv170200786A} may be necessary for detection and study of such systems. 
\item {\it Newtonian and Relativistic Precession}: Our analysis did not consider the effects of Newtonian precession or relativistic precession in compact binary systems. These effects can introduce additional complexities into the observed variability, and further investigation is required to assess their significance \citep[see the example of OJ287][]{Laine_2020}.
\item {\it Contribution of Circumbinary Structures}: The potential contribution of circumbinary structures, such as circumbinary disks or circumbinary broad line regions, to the emission from CB-SMBH systems could impact the observed photocenter motion. The extent of this impact and its implications for binary framework studies remain subjects for future exploration \citep{Dexter_2020, 2022A&A...663A..99K, 2021MNRAS.505.5192P, 2022AN....34310073S, 2022arXiv220706432G}.
\end{itemize}

\section{Conclusions}\label{sec:concl}
We  developed a  two-tier Bayesian framework for the {synthesis} of  multi-instrument observational data from CB-SMBH systems. The Simulator, as the first tier, is  generating synthetic datasets that replicate the astrometric wobble and total optical light curves of these systems. The second tier, the Solver, uses these datasets to infer individual fluxes of binary components and binary orbital parameters with reliable precision, as evidenced by our simulation's posterior distributions.

In summary, these are our conclusions:
\begin{itemize}

\item
{
 Our  Bayesian synthesis method reliably  determines binary orbital parameters and extract individual fluxes of binary components from the total flux, within multi, intermediate and short-term  observational cycles with respect to orbital period, assuming density of observations 10 per year in astrometry and photometry.
 For configurations with longer periods, lower eccentricity and mass ratios, the method accurately tracks component fluxes within 1 HDI. In configurations with higher eccentricity and mass ratio over a one-year period, it demonstrates rapid dynamic analysis capabilities, also achieving  decompositions of fluxes within 1 HDI. Configurations with higher eccentricities and intermediate obsevrational baselines ($L_{oc}\rightarrow{1}$)
  result in decompositions of components fluxes within 2 HDI. Case with short period, high eccentricity, and significant mass, particularly in "q-accrete" scenario with $L_{oc}>1$, show the method's robustness in capturing swift variations, with decompositions of fluxes of components aligning within 3 HDI.}

\item
{
The configurations analyzed highlight the need for adaptable observational strategies to accommodate diverse binary system characteristics. Key considerations include beginning with a standard observation rate (e.g. 10 observations per year) but remaining flexible to adjust based on initial data analysis. Increased frequency may be used for systems showing rapid flux variations or complex orbital dynamics. Using multi-wavelength observations can provide insights into the binary system's nature, informing adjustments to observational strategies for better targeting observed features. An illustrative workflow might involve leveraging LSST catalogue data on variability to inform real-time optimization of observing strategies by future astrometry facilities.}

\item
{The astrometric} wobble is particularly sensitive to factors such as the semi-major axis, eccentricity, and orbital phase, with the largest values  occurring in highly eccentric orbits at the apocenter.
Association between the binary mass ratio and total photocenter offset, suggests that the more massive component's contribution to the system's total light curve is inversely related to its luminosity. This trend is especially pronounced in systems with total masses  $> 10^8 M_{\sun}$ and periods of $\geq$ 1 year, where larger masses and longer periods typically result in greater photocenter displacements. We show that this can provide  preliminary information on components' activity.
\end{itemize}

Our  Bayesian approach shows the potency of multi-instrument data integration and allows for CB-SMBH observational timelines  over 1-2 orbital periods, even for  systems with periods $\rightarrow 1$ year. Such a synthesis can assist in extracting information from upcoming large-scale time-domain surveys, like LSST, and next-generation interferometers, such as ng-EHT and GRAVITY+.

\begin{acknowledgments}
ABK and L{\v C}P acknowledge funding provided by University of Belgrade-Faculty of Mathematics  (the contract \textnumero{451-03-66/2024-03/200104}), and Astronomical observatory Belgrade (the contract \textnumero{451-03-66/2024-03/200002} through the grants by the Ministry of Science, Technological Development and Innovation   of the Republic of Serbia.
\end{acknowledgments}

%% To help institutions obtain information on the effectiveness of their 
%% telescopes the AAS Journals has created a group of keywords for telescope 
%% facilities.
%
%% Following the acknowledgments section, use the following syntax and the
%% \facility{} or \facilities{} macros to list the keywords of facilities used 
%% in the research for the paper.  Each keyword is check against the master 
%% list during copy editing.  Individual instruments can be provided in 
%% parentheses, after the keyword, but they are not verified.

%\vspace{5mm}
%\facilities{HST(STIS), Swift(XRT and UVOT), %AAVSO, CTIO:1.3m,
%CTIO:1.5m,CXO}

%% Similar to \facility{}, there is the optional \software command to allow 
%% authors a place to specify which programs were used during the creation of 
%% the manuscript. Authors should list each code and include either a
%% citation or url to the code inside ()s when available.
%%%%%%%%%%%%%%%%SOFTWARE
\software{PyMC \citep{Salvatier2016, pymc_2022}, pymc\_ext \citep{exoplanet_joss}
}

%% Appendix material should be preceded with a single \appendix command.
%% There should be a \section command for each appendix. Mark appendix
%% subsections with the same markup you use in the main body of the paper.

%% Each Appendix (indicated with \section) will be lettered A, B, C, etc.
%% The equation counter will reset when it encounters the \appendix
%% command and will number appendix equations (A1), (A2), etc. The
%% Figure and Table counter will not reset.

\appendix

\section{Number of CB-SMBH with astrometric wobble }
\label{sec:apendno}

{At angular-diameter distance $D(z)$, and $LOS=i$ the binary orbital projected angular extent is $\theta \sim   \left( \frac{G P^2 (M_{tot})}{4 \pi^2} \right)^{\frac{1}{3}}
 \sin i/D(z)$.
The  astrometric wobble above the given threshold $o$ is 
$|\delta_{ph}|=|\theta \frac{1+(q/l)}{1-(q/l)}|>o$. 
 Then, instrument  can detect astrometric wobble if 
%based on equation that wobble\sim \theta(1-q/l)(1+q/l)
$\mathcal{H}(\theta,o,q,l): \theta>|o\frac{1+(q/l)}{1-{q/l)}}|$  and if the orbital period $\leq$  the half of  instrument observational timeline.
We use the quasar luminosity function $\frac{d^{2}N}{d\log LdV}$ \citep{Hopkins_2007, 10.1093/mnras/stab1856}to derive
the number of AGN per redshift $z$ and luminosity $L$.At each
total binary mass  bin we derive luminosity from the assumption that the AGN emits at a fraction of Eddington luminosity, $L=f_{Edd}L_{Edd}M_{tot}$, where for the  Eddington fraction of bright AGN we adopt $f_{Edd}=0.1$. We
assume that quasars typically have a total lifetime of $t_{Q} \sim  \text{few} \times 10^{7}$
years (independent of redshift and luminosity), and that their residence time at given $P,q$ is
\begin{equation}
t_{res}(P,M,q)= \frac{a}{\dot{a}} = \frac{20}{256}\left( \frac{P}{2\pi} \right)^{\frac{8}{3}} \left( \frac{GM_{tot}}{c^3} \right)^{-\frac{5}{3}} \left(\frac{4q}{(1+q)^{2}}\right)^{-1}.
\end{equation}}

{Under these assumptions, the
number of quasars with orbital periods $P=[0.5, 1]$ year, and $q=[0.67,1], l=[0.1,1]$ is given approximately by
\begin{equation}
N_{\text{SBHB}} = f_{\text{bin}} \int_{0}^{\infty} \left\{ 4\pi \int \frac{d^{2}V}{dzd\Omega} \int_{\log L_{\text{min}}(z)}^{\infty} \frac{d^{2}N}{d\log LdV} \frac{t_{res}(P,M,q)}{t_{Q}} \times \mathcal{H}(\theta,o,q,l) d\log L \right\} dz
\end{equation}
The differential comoving volume element is calculated for cosmological parameters 
$H_{0}=70\, \mathrm{km s}^{-1}\mathrm{Mpc}^{-1}$, $\Omega_{\Lambda}=0.7$,
$\Omega_{m}=0.3$ \citep{Planck_2020} via \texttt{astropy.cosmology} module.}

{With this approximation, we compute the distribution of the number of CB-SMBH in
a grid $50\times 50$ of mass bins evenly distributed across $10^{6-10}M_{\odot}$ in log-space, and redshift bins in $[0.01, 0.03]$, respectively.}

\section{Simulator  of CB-SMBH astrometric data and light curve }
\label{sec:apend1}

\subsection{Simulation of CB-SMBH Astrometric Data}\label{sec:astrom}

The CB-SMBH is assumed {to be a} two-body system of SMBHs,  { such that their masses satisfy $M_{1}>M_{2}$}. 
The procedure for simulating astrometric data  is {discussed} briefly {below}, and further {information} may be found in  \citet{kovacevic2020}. 

\begin{table}
	\caption{Model parameters of a Keplerian orbit of {CB-SMBH} in three dimensions.}
	\centering
	\setlength{\tabcolsep}{1pt}
	\begin{tabular}{lccc}
		\hline\hline
{Parameter} & {Units}&{Name}& {fiducial range}\\  
		\hline
$a$ & ld $\vee$ pc&semimajor axis & $[0,\infty)$ \\
$e$ & -&eccentricity &$[0,1]$ \\
$P$ & yr $\vee$ days&orbital period &$(0,+\infty)$ \\
$\omega $& $^{\circ}$&argument of pericenter&$[0, 360]$\\
$i $& $^{\circ}$&inclination&$[-90, 90]$\\
$\Omega $& $^{\circ}$&angle of ascending node&$[0, 360]$\\
$T_{0}$ & days $\vee$ yr&time of pericenter passage & $[0,\infty)$\\
			\hline
	\end{tabular}
	\label{deforbit}
\end{table}

{Naturally, dynamical parameters fully describe the SMBH position relative to the barycentre of system (see Table \ref{deforbit}).
	The apparent relative orbit is the one of the secondary around the primary projected on the sky plane, and it can be determined from measurements of the relative position of the components obtained through astrometric imaging or interferometric observations.
	The projected spatial motion of the binary components is calculated within the reference frame centered on the primary component or barycentre and two axes in the plane tangent to the celestial sphere \citep{2013A&A...551A.121L}:  the $x$-axis points north, while the $y$-axis points east. The $z$-axis runs parallel to the line of sight and points in the direction of rising radial velocities (positive radial velocities). Here we provide equation in the barycenter reference frame.}

The transformations could be {represented} in the vectors $\mathbf{P}$ and $\mathbf{Q}$ (Thiele-Innes parameters) instead of {utilizing} cosine and sine terms of Euler rotations. 
 
The vector of relative position $\mathbf{s}(t)=[s_{x}(t),s_{y}(t),s_{z}(t)]$ of an SMBH 
w.r.t barycentre  is:
\begin{equation}
\mathbf{s}(t)=\mathbf{s(0)}+\partial_{t}\mathbf{s}(0)t +\mathbf{P}[\cos E(t)-e]+\mathbf{Q}\sqrt{1-e^2}\sin E(t),
\label{eq:compose}
\end{equation}
where $E(t)$ is {the }eccentric anomaly {determined} from the Kepler equation $E(t)-e\sin E(t)=2\pi n (t-t_{0}), n=P^{-1}$, {and} $t_{0}$ {is set to} $0$  for simplicity. 
The vectors of barycenter position 
$\mathbf{s}(0)$ and velocity $\partial_{t}\mathbf{s}(0)$ are set to zero for simplicity. Thiele-Innes functions $\mathbf{P}$ and $\mathbf{Q}$ are defined as follows:
\begin{eqnarray}\label{thiel1}
\mathbf{P} &=& a \left[\mathbf{p}\cos(\omega)+\mathbf{q} \sin(\omega)\right],\\
\label{thiel2}
\mathbf{Q} &=& a \left[-\mathbf{p}\sin(\omega)\ + \mathbf{q}\cos(\omega)\right],\\
\label{thiel3}
\mathbf{p}&=&(\sin \Omega,\cos \Omega,0),\\
\label{thiel4}
\mathbf{q} & =&\left(\mathcal{I}\cos{\Omega},-\mathcal{I}\sin\Omega,\sin i\right).%\\
%\label{thiel5}
\end{eqnarray}
\noindent where we  assume that $a=M_{1}G^{1/3}(n\cdot M_{tot})^{1/3}$ is semimajor axis of secondary component, and $I=\cos i$. Therefore, above equations describe relative orbit of secondary with respect to primary.
The data matching {the third coordinate of body position} ( $s_{z}(t)$) can not be {obtained}.

The model for photocenter motion is now simply
$$\Theta(t)=[\Theta_{x}(t),\Theta_{y}(t)]=D^{-1}(\mu-l(t))[s_{x}(t),s_{y}(t)].
$$
\noindent where $D$ is a distance,  $q=M_{2}/M_{1}$, $f(t)=F_{2}/F_{1}$, $ \mu=\frac{M_{2}}{M_{1}+M_{2}}=\frac{q}{1+q}$ and $l(t)=\frac{F_{2}}{F_{1}+F_{2}}=\frac{f(t)}{1+f(t)}$.

 We assume that both masses, and distance are known and that fluxes of both SMBH are obtained via simulation procedure given in Appendix \ref{sec:stochastic}.
We assume that the errors for each artificial observation are independent and identically distributed, resembling white noise at the level of $5\%$ \citep[see also][]{Dexter_2020}. 
{In order to avoid using} the same model for both {producing} the observations and finding the  solution, an additional jitter was {added} in the model when {simulating} the data.  
 \subsubsection{Approximation of Photocenter Motion Using True Anomaly}
In this section, we will re-approximate the motion of the photocenter using the true anomaly ($\nu$).  It represents the angle between the direction of periapsis (the point of closest approach to the central body) and the current position of the object. As such, it is a more intuitive parameter that can be visualized and understood easily.
More importantly, the true anomaly proves particularly useful for predicting an object's future position on its orbit. This property is advantageous for the general exploration of the parameter grid space of CB-SMBH. The use of true anomaly allows for a more straightforward and less computationally-intensive approximation.

Let the two components of binary are at the relative position given by $(\bm{\xi}_{2}-\bm{\xi}_{1})=\frac{a(1-e^{2})}{1+e\cos\nu}\mathcal{P}^{T}(i,\omega,\Omega) (\cos\nu,sin\nu,0)$, where $^{T}$ mark transposition, $\mathcal{P}$ is orbit orientation matrix of binary, and $\frac{a(1-e^{2})}{1+e\cos\nu} (\cos\nu,\sin\nu,0)$ is a polar coordinates in the orbital plane as given in Appendix A in \citet{kovacevic2020}.
We then rewrite the components $s_{x}, s_{y}$ given in   Eq. \ref{eq:compose} in terms of orbital phase $\nu$:
\begin{align}
s_{x}= r ( \cos(\Omega) \cos(\nu +\omega) - \sin(\Omega) \sin(\nu + \omega) \cos i )\\
s_{y} = r ( \sin(\Omega) \cos(\nu + \omega) + \cos(\Omega) \sin(\nu + \omega) \cos i )\\
r=\frac{a(1-e^{2})}{1+e\cos\nu}
\end{align}
Then we will calculate  the total photocenter motion as :
\begin{align}
  \lbrack\bm{\delta}_{ph}\rbrack &
    =(\mu-l(t))  \sqrt{s^{2}_{x}+s^{2}_{y}} \\
    &\sim (\mu-l(t))r \sqrt{ 1 - \sin^{2}(\nu+\omega) sin^{2} i 
}
\label{eq:photonewx}
\end{align}

\subsection{Simulation of CB-SMBH  light curve}\label{sec:emission}

At higher viewing angles, the detected binary
radiation would be strongly modulated by relativistic effects
such as gravitational lensing and Doppler boosting. Here, we consider the case when  the emission coming from two minidiscs surrounding each component in the system will appear Doppler boosted \citep{Haiman_2009, 10.1093/mnras/staa1312}.
We schematically decompose the total observed flux $\tilde{F}_{tot}$ of binary  as  \citep[see, e.g.][]{10.1093/mnras/staa1312,Guti_rrez_2022}

\begin{align}
    \tilde{F}_{tot}&=\bar{F}_{1}(1+\Delta F_{\mathcal{D}1}) +F_{1}+ \bar{F}_{2}(1+\Delta F_{\mathcal{D}2}) +F_{2}\\
    &=(\bar{F}_{1}+\bar{F}_{2})(1+\Delta F_{\mathcal{D}1}\frac{\bar{F}_{1}}{\bar{F}_{1}+\bar{F}_{2}}+\Delta F_{\mathcal{D}2}\frac{\bar{F}_{2}}{\bar{F}_{1}+\bar{F}_{2}})+F_{1}+F_{2}
    \label{eq:generalfluxmodel22}
\end{align}
\noindent where the  $F_{i}$, $\bar{F}_{i}, \Delta F_{\mathcal{D}i}, \, i=1,2$ account for stochastic variability, corresponding average values and Doppler boost term of each minidisc in the system, respectively. 

The next two sections provide formulas for  Doppler brightnening and stochastic variability calculation for each minidisk. 

\subsubsection{Doppler boost}\label{sec:doppler}
At a given photon frequency $\nu_p$, assuming that the minidisc emission is  $F_{i}\propto \nu_{p}^{\alpha}$
the  Doppler effect is
\citep{2018MNRAS.474.2975D}
\begin{equation}
\Delta F_{\mathcal{D}{i}}=(3-\alpha_{\nu_{p}})(v_{i}/c) \cos \phi \sin I , i=1,2
\end{equation}
\noindent  where $\alpha_{\nu_{p}}$
$I$, $\phi$, and $v_{i}$ are the power law spectrum index, orbital inclination, phase of binary, and 3D velocity of component, respectively. 
The projection of the velocity vector onto the line of sight is given as

\begin{align}
    v_{i}&=K_{i}(\cos(\omega + \nu) +e \cos(\omega), i=1,2\\
    K_{1}&=\frac{2\pi}{P} \frac{q}{1+q} \frac{a \sin I}{\sqrt{1-e^2}}\\
    K_{2}&=\frac{2\pi}{P} \frac{1}{1+q} \frac{a \sin I}{\sqrt{1-e^2}}
\end{align}
\noindent $\omega$ is the argument of pericenter, and $\nu$ is the
true anomaly.
We solve for the radial velocity for both binary components, prescription given in \citet{2022A&A...663A..99K, 10.1051/0004-6361/202038733} for binary mass ratio $q < 1$.
Both components are assumed to be Doppler-modulated with corresponding line-of-sight velocities and that emission from
both black holes share the same spectral index with a value  $\alpha_{\nu_{p}}\sim -0.44$ of the composite quasar spectrum \citep{2001AJ....122..549V}.

\subsubsection{Stochastic variability}\label{sec:stochastic}

It is predicted that the quasars high-quality light curves obtained from the LSST and other large time domain facilities will be not only shifted in time but also distorted at different wavelengths \citep{2020A&A...636A..52C,2022ApJS..262...49K}.  We adopted this approach to model the stochastic emission $F_{i}$ of each minidisk in the context of the lamp-post model, thin-disk geometry,and damped random walk (DRW) for the driving function, where the size of the transfer function expands with wavelength.

For this purposes we will introduce auxiliary scaling factors.  To approximate the mass accretion rate  $\dot{M}_{tot}$ of the binary system in terms of the total mass of the binary $M_{tot}$,  we can use the Eddington accretion rate:
\begin{equation}
  \dot{M}_{tot}=4 \pi m_{p} G M_{tot}/{\epsilon c \sigma_{T}}\propto \rho M_{tot}
\end{equation} 
\noindent  where $m_p$ is the proton mass, $\sigma_T$ is the Thompson
scattering cross-section, and $\epsilon$ is a typical radiative efficiency of quasar, and $\rho$ is a constant. 
For the symmetrical binary systems $q=1$, it is expected that $\dot{M}_{1}=\dot{M}_{2}$, and as the $q$ decreases, the  ratio $\eta=\dot{M}_{2}/\dot{M}_{1}$ increases \citep{10.1111/j.1365-2966.2009.14796.x, 2009ApJ...690...20S, 2020ApJ...901...25D}.
Therefore, if the implied accretion rate of the binary  is $\dot{M}_{tot}=\dot{M}_{1}+\dot{M}_{2}$,   then the accretion rate is split between the two black holes  with
  $\dot{M}_{i}= f_{i} M_{tot}, i=1,2 $ where $f_{1}=(1+\eta)^{-1} \rho$ and  $f_{2}=(1+\eta^{-1})^{-1}\times\rho$.
{ We  assume that $\eta=(1+0.9q)^{-1}$ as given in \citet{2020ApJ...901...25D}, if not otherwise stated}. 
Each minidisk in a binary is set up  as a non-relativistic thin-disk  emitting black-body radiation \citep{1973A&A....24..337S, 1974ApJ...191..507T}, with the central source treated as a "lamp post" placed near the black hole \citep{2007MNRAS.380..669C, 2017ApJ...835...65S}.
Then, the temperature profiles  $T_{i}, i=1,2$ of each minidisk  are described as \citep{2020A&A...636A..52C,2022ApJS..262...49K,2017ApJ...835...65S, 2007MNRAS.380..669C, PhysRevLett.128.191101}

\begin{align}
    T_{i}\propto \Bigg(\left(\frac{r} {R_{0}}\right)^{-3}\Big(1-\sqrt{\frac{r_0}{r}}\Big)\Bigg)^{1/4} 
    \label{eq:temp}
\end{align}
\noindent where the scale radius is

\begin{equation}
    R_{0}=\Bigg(\left(\frac{k \lambda}{h c}\right)^{4}\frac{ 3 G M_{i} \dot{M}_{i}}{8\pi \sigma}\Bigg)^{1/3}=\Bigg(\left(\frac{k \lambda}{h c}\right)^{4}\frac{ 3 G M_{tot}^{2} f_{i} \tilde{q}}{8\pi \sigma}\Bigg)^{1/3},
    \label{eq:radius0}
\end{equation}
\noindent $\lambda$ is the rest frame  wavelength, $\sigma$ is the Stefan-Boltzmann constant, $k$ is Boltzmann constant, $G$ is a gravitational constant, $M_{tot}$ is total mass of the binary, $r_{0}=6 G M_{i}/c^{2}$ is the inner radius of minidisk surrounding mass $M_i$ and  
\begin{equation}
    \tilde{q}=
  \begin{cases}
  \frac{1}{1+q} & \text{for} M_{1} \\
 \frac{q}{1+q} &  \text{for} M_{2}
  \end{cases}
  \label{eq:cases}
\end{equation}

Using Eqs. \ref{eq:temp}-\ref{eq:cases} and following prescription in \citet{2020A&A...636A..52C,2022ApJS..262...49K}, the transfer function for each disc at given $\lambda$  is:

\begin{align}
 \psi_{i} (\tau, \lambda)\propto & \int^{r_{out}}_{r_{0}} \int^{2\pi}_{0}\frac{\zeta e^{\zeta}}{(e^{\zeta}-1)^{2}}  r dr d\theta \cos i \times \\
 & \delta(\tau -(r/c)(1-\sin i \cos \theta))\nonumber \\
 \zeta=\frac{hc}{k\lambda T_{i}}
 \label{eq:transfer}
\end{align}
\noindent where the outer disc radius is approximated as $r_{out}\sim 0.5 r_{0} (M_{i})^{2/3}$ \citep{2014ApJ...783...47J},   $\theta$ is azimuth of reprocessing site, $i$ is inclination of the disc.

Finally, assuming that each minidisk has an own driving-source
emission  $\mathcal{F}_{i}, i=1,2$,  the total minidisc flux $F_{i}(\lambda,t), i=1,2$ is obtained by convolution  \citet{2018MNRAS.473...80T}:

\begin{equation}
    F_{i}(\lambda,t)\sim \sum^{\tau_{max}}_{\tau=\tau_{0}} \psi_{i}(\tau,\lambda)\mathcal{F}_{i}(t-\tau)\Delta \tau
    \label{eq:convl}
\end{equation}

Here, $\mathcal{F}_{i}(t-\tau)$ is the fractional luminosity variability `lagged' by the light travel time
$\tau=r/c$ from the disc center.
We represent this driving light curve by the Damped Random Walk (DRW) model with the characteristic amplitude $\sigma_{DRW}$ and timescale $\tilde{\tau}$  \citep{Kelly2009, Kozlowski2017}.
The DRW is constructed from the procedure given in \citet{Kovacevic2021}.

\section{Additive decomposition of CB-SMBH light curves using GP} %\appendix
\label{sec:sectionB0}
Here we provide an auxiliary formula for  Gaussian conditional (Section \ref{sec:sectionB}), which enables us to model the conditional distribution of one GP component given the sum of GPs. Then we proceed with the likelihood of additive  decomposition of CB-SMBH light curves via GPs (Section \ref{sec:GP}), which provides a means to estimate the individual components from the sum of the light curves of both SMBH in binary system.

\subsection{Gaussian Conditional} %\appendix
\label{sec:sectionB}

Here we provide an auxiliary  formula for the multivariate Gaussian conditional distribution $\mathbf{x}_{A} | \mathbf{x}_{B}$, 
which  is used  for likelihood of  additive decomposition of Gaussian process \citep[see][]{Rasmussen2004,  mardia79, duvenaud13}.

Let $x$ follow a multivariate normal distribution:

\begin{equation} 
x \sim \mathcal{N}(\mu, \Sigma).
\label{eq:mvn}
\end{equation}

Then, the conditional distribution of any subset vector $x_A$, given the complement vector $x_B$, is also a multivariate normal distribution:

\begin{equation} 
x_A|x_B \sim \mathcal{N}(\mu_{A|B}, \Sigma_{A|B})
\label{eq:mvn-cond}
\end{equation}

where the conditional mean and covariance are:

\begin{equation} 
\begin{aligned}
\mu_{A|B} &= \mu_A + \Sigma_{AB} \Sigma_{BB}^{-1} (x_B - \mu_B) \\
\Sigma_{A|B} &= \Sigma_{AA} - \Sigma_{AB} \Sigma_{BB}^{-1} \Sigma_{BA}
\end{aligned}
\label{eq:mvn-cond-hyp}
\end{equation}

with block-wise mean and covariance defined as:

\begin{equation} 
\begin{aligned}
\mu &= \begin{bmatrix} \mu_A \\ \mu_B \end{bmatrix} \\
\Sigma &= \begin{bmatrix} \Sigma_{AA} & \Sigma_{AB} \\ \Sigma_{BA} & \Sigma_{BB} \end{bmatrix} .
\end{aligned}
\label{eq:mvn-joint-hyp}
\end{equation}

To prove the concept of \ref{eq:mvn-cond-hyp}, we will assume that \ref{eq:mvn-joint-hyp} and following equation holds:
\begin{equation}
x = \begin{bmatrix} x_A \\ x_B \end{bmatrix}
 \label{eq:x}
\end{equation}
where $x_A$ is an $n_A \times 1$ vector, $x_B$ is an $n_B \times 1$ vector, and $x$ is an $n_A + n_B = n \times 1$ vector.
Then, we recall that  by construction, the joint distribution of $x_A$ and $x_B$ is:

\begin{equation} 
x_A,x_B \sim \mathcal{N}(\mu, \Sigma).
\label{eq:mvn-joint}
\end{equation}

Moreover, the marginal distribution of $x_B$ follows from a multivariate normal distribution \ref{eq:mvn} and \ref{eq:mvn-joint-hyp} as
\begin{equation} 
x_B \sim \mathcal{N}(\mu_B, \Sigma_{BB}).
\label{eq:mvn-marg}
\end{equation}

According to the Bayes law of conditional probability, it holds that
\begin{equation} 
p(x_A|x_B) = \frac{p(x_A,x_B)}{p(x_B)}
\label{eq:mvn-cond-s1}
\end{equation}

Applying \ref{eq:mvn-joint} and \ref{eq:mvn-marg} to \ref{eq:mvn-cond-s1}, we have:
\begin{equation} 
p(x_A|x_B) = \frac{\mathcal{N}(x; \mu, \Sigma)}{\mathcal{N}(x_B; \mu_B, \Sigma_{BB})}.
 \label{eq:mvn-cond-s2}
\end{equation}

Using the probability density function of the multivariate normal distribution, we get:

\begin{equation}\label{mvn-cond-s3}
\begin{split}
p(x_A|x_B) &= \frac{\frac{1}{\sqrt{(2 \pi)^n |\Sigma|}} \cdot \exp \Bigg[ -\frac{1}{2} (x-\mu)^\mathrm{T} \Sigma^{-1} (x-\mu) \Bigg]} 
{\frac{1}{\sqrt{(2 \pi)^{n_B} |\Sigma_{BB}|}} \cdot \exp \Bigg[ -\frac{1}{2} (x_B-\mu_B)^\mathrm{T} \Sigma_{BB}^{-1} (x_B-\mu_B) \Bigg] } =\\
&\frac{1}{\sqrt{(2 \pi)^{n-n_B}}} \cdot \sqrt{\frac{|\Sigma_{BB}|}{|\Sigma|}} \cdot \exp \Bigg[ -\frac{1}{2} (x-\mu)^\mathrm{T} \Sigma^{-1} (x-\mu) + \\
&\frac{1}{2} (x_B-\mu_B)^\mathrm{T} \Sigma_{BB}^{-1} (x_B-\mu_B) \Bigg] .
\end{split}
\end{equation}

Writing the inverse of $\Sigma$ as

\begin{equation}
\Sigma^{-1} = \begin{bmatrix} \Sigma^{AA} & \Sigma^{AB} \\ \Sigma^{BA} & \Sigma^{BB} \end{bmatrix}
 \label{eq:Sigma-inv-def}
\end{equation}

and applying \ref{eq:mvn-joint-hyp} to \ref{mvn-cond-s3}, we get:

\begin{equation} 
\begin{split}
p(x_A|x_B) = &\frac{1}{\sqrt{(2 \pi)^{n-n_B}}} \cdot \sqrt{\frac{|\Sigma_{BB}|}{|\Sigma|}} \cdot \\
&\exp \Bigg[ -\frac{1}{2} \Bigg( \begin{bmatrix} x_A \\ x_B \end{bmatrix} - \begin{bmatrix} \mu_A \\ \mu_B \end{bmatrix} \Bigg)^\mathrm{T} \begin{bmatrix} \Sigma^{AA} & \Sigma^{AB} \\ \Sigma^{BA} & \Sigma^{BB} \end{bmatrix} \Bigg( \begin{bmatrix} x_A \\ x_B \end{bmatrix} - \begin{bmatrix} \mu_A \\ \mu_B \end{bmatrix} \Bigg) \\
&\exp \Bigg[ + \frac{1}{2} \, (x_B-\mu_B)^\mathrm{T} \, \Sigma_{BB}^{-1} \, (x_B-\mu_B) \Bigg] .
\end{split}
\label{eq:mvn-cond-s44}
\end{equation}

Rearranging  \ref{eq:mvn-cond-s44}, we have

\begin{equation} 
\begin{split}
p(x_A|x_B) = &\frac{1}{\sqrt{(2 \pi)^{n-n_B}}} \cdot \sqrt{\frac{|\Sigma_{BB}|}{|\Sigma|}} \cdot \\
&\exp \Bigg[ -\frac{1}{2} \Bigg( (x_A-\mu_A)^\mathrm{T} \Sigma^{AA} (x_A-\mu_A) + \\
&2 (x_A-\mu_A)^\mathrm{T} \Sigma^{AB} (x_B-\mu_B) + (x_B-\mu_B)^\mathrm{T} \Sigma^{BB} (x_B-\mu_B) \\
&\hphantom{\exp \Bigg[ -\frac{1}{2} \Bigg(} + \frac{1}{2} (x_B-\mu_B)^\mathrm{T} \Sigma_{BB}^{-1} (x_B-\mu_B) \Bigg] \Bigg].
\end{split}
\label{eq:mvn-cond-s5}
\end{equation}

Based on the inverse of a block matrix, the inverse of $\Sigma$ in \ref{eq:Sigma-inv-def} is:
\begin{align}\label{eq:inverse}
\small
\begin{bmatrix}
    \Sigma_{AA} & \Sigma_{AB} \\
    \Sigma_{BA} & \Sigma_{BB}
\end{bmatrix}^{-1} &= 
 \begin{bmatrix}
    X & Y \\
    Z & W
\end{bmatrix},
\end{align}
\noindent where block matricies are:
\begin{align*}
X &= (\Sigma_{AA} - \Sigma_{AB} \Sigma_{BB}^{-1} \Sigma_{BA})^{-1}, \\
Y &= -(\Sigma_{AA} - \Sigma_{AB} \Sigma_{BB}^{-1} \Sigma_{BA})^{-1} \Sigma_{AB} \Sigma_{BB}^{-1}, \\
Z &= -\Sigma_{BB}^{-1} \Sigma_{BA} (\Sigma_{AA} - \Sigma_{AB} \Sigma_{BB}^{-1} \Sigma_{BA})^{-1}, \\
W &= \Sigma_{BB}^{-1} + \Sigma_{BB}^{-1} \Sigma_{BA} (\Sigma_{AA} - \Sigma_{AB} \Sigma_{BB}^{-1} \Sigma_{BA})^{-1} \Sigma_{AB} \Sigma_{BB}^{-1}.
\end{align*}

We also recalll that the determinant of a block matrix is:
\begin{equation}\label{eq:determinant}
\begin{vmatrix}
    \Sigma_{AA} & \Sigma_{AB} \\
    \Sigma_{BA} & \Sigma_{BB}
\end{vmatrix} = |\Sigma_{BB}||\Sigma_{AA}-\Sigma_{AB}\Sigma^{-1}_{BB}\Sigma_{BA}|.
\end{equation}

Plugging \ref{eq:inverse} and \ref{eq:determinant} into \eqref{eq:mvn-cond-s5} and doing some algebra operations, we have:

\begin{equation}
\begin{split}
p(x_A|x_B) = &\frac{1}{\sqrt{(2 \pi)^{n-n_B}}} \cdot \sqrt{\frac{|\Sigma_{BB}|}{|\Sigma|}} \cdot \\
&\exp \Bigg[ -\frac{1}{2} \Bigg[ (x_A-\mu_A)^\mathrm{T} M^{-1} (x_A-\mu_A) \\
&\hphantom{\exp \Bigg[ -\frac{1}{2} \Bigg[} - 2 (x_A-\mu_A)^\mathrm{T} M^{-1} \Sigma_{AB} \Sigma_{BB}^{-1} (x_B-\mu_B) \\
&\hphantom{\exp \Bigg[ -\frac{1}{2} \Bigg[} + (x_B-\mu_B)^\mathrm{T} N (x_B-\mu_B) \Bigg] \Bigg].
\end{split}
\label{eq:mvn-cond-s8}
\end{equation}

\noindent where

\begin{align*}
M &= \Sigma_{AA} - \Sigma_{AB} \Sigma_{BB}^{-1} \Sigma_{BA}, \\
N &= \Sigma_{BB}^{-1} + \Sigma_{BB}^{-1} \Sigma_{BA} M^{-1} \Sigma_{AB} \Sigma_{BB}^{-1}.
\end{align*}

With $\Sigma_{BA} = \Sigma_{AB}^\mathrm{T}$, because $\Sigma$ is a covariance matrix, and $n - n_B = n_A$, we finally arrive at

\begin{equation}
\begin{split}
p(x_A|x_B) = &\frac{1}{\sqrt{(2 \pi)^{n_A} |M|}} \\
&\exp \Bigg[ -\frac{1}{2} \Bigg[ (x_A - \mu_A)^\mathrm{T} M^{-1} (x_A - \mu_A) \\
&\hphantom{\exp \Bigg[ -\frac{1}{2} \Bigg[} - 2 (x_A - \mu_A)^\mathrm{T} M^{-1} \Sigma_{AB} \Sigma_{BB}^{-1} (x_B - \mu_B) \Bigg] \Bigg].
\end{split}
\label{eq:mvn-cond-s9}
\end{equation}

The above equation is the probability density function of a conditional  multivariate normal distribution:

\begin{equation}
p(x_A|x_B) = \mathcal{N}(x_A; \mu_{A|B}, \Sigma_{A|B})
\label{eq:mvn-cond-s10}
\end{equation}

with the mean $\mu_{A \vert B}$ and covariance $\Sigma_{A \vert B}$ given by \eqref{eq:mvn-cond-hyp}.

\subsection{Likelihood of Decomposition of CB-SMBH Light Curve via GP
}\label{sec:GP}

Let the CB-SMBH observed light curve  is $F_{tot}(t)=F_{1}(t)+F_{2}(t)$, and  a priori unknown and independent  emissions from components are given as $F_{i}(t), i=1,2$ at observed time instances $t={e}_{1},..{e}_{n}$.
We assume that  the light curves have the following Gaussian Process representation \citep{Rasmussen2004}:

\begin{align}
    F_{1}(t) &\sim G(\mu_{1}(t),k_{1}(t,t))\\
    F_{2}(t) &\sim G(\mu_{2}(t),k_{2}(t,t)\\
    F_{tot}(t)&=F_{1}(t)+ F_{2}(t) \sim G(\mu_{1}(t)+\mu_{2}(t),k_{1}(t,t)+k_{2}(t,t))\\
    \label{eq:gpgeneral}
\end{align}

\noindent where $G$ is Gaussian distribution with mean function $\mathbb{E}(F_{i}(t))=\mu_{i}(t), i=1,2$ and covariance functions $Cov(F_{i}(t),F_{i}(t')= k_{i}(t,t')$  (also called the kernels). 
Any number of components in multiple systems can be summed this way.
In what follows, we distinguish between  flux values
at training locations $t={e}_{1},..{e}_{n}$ and the flux values at some set
of query locations $t^{\star}={e}^{\star}_{1},..{e}^{\star}_{n}$

Then, the joint prior distribution over two unknown component light curve drawn independently
from GP priors, and their sum is given by \citep{duvenaud13}:

\begin{align}
\begin{split}
      \begin{bmatrix}
           F_{1}(t) \\
           F_{1}(t^{\star}) \\
            F_{2}(t) \\
           F_{2}(t^{\star} )\\
          F_{1}(t)+ F_{2}(t) \\
          F_{1}(t^{\star})+  F_{2}(t^{\star})
         \end{bmatrix} \sim \\
         \end{split}\\
         \begin{split}
         G\begin{pmatrix}
         \begin{bmatrix}
           \mu_{1} \\
           \mu^{\star}_{1} \\
           \mu_{2} \\
   \mu^{\star}_{2} \\
              \mu_{1} + \mu_{2}\\
              \mu^{\star}_{1} + \mu^{\star}_{2} 
 \end{bmatrix},
  \begin{bmatrix}
 \kappa_{1} & \kappa^{\star}_{1} & 0 & 0&\kappa_{1} & \kappa^{\star}_{1}\\ 
 \kappa^{{\star} T}_{1} & \kappa^{{\star}{\star}}_{1} & 0 & 0&\kappa^{\star}_{1} & \kappa^{{\star}{\star}}_{1}\\ 
  0 & 0&\kappa_{2} & \kappa^{\star}_{2} & \kappa_{2} & \kappa^{\star}_{2} \\ 
   0 & 0& \kappa^{{\star}{T}}_{2} & \kappa^{{\star}{\star}}_{2}  &  \kappa^{\star}_{2} & \kappa^{{\star}{\star}}_{2}\nonumber\\ 
    \kappa_{1} & \kappa^{{\star}{T}}_{1} & \kappa_{2} & \kappa^{{\star}{T}}_{2} &
    \kappa_{1}+\kappa_{2}&\kappa^{\star}_{1}+ \kappa^{\star}_{2}\\
     \kappa^{{\star}{T}}_{1} & \kappa^{{\star}{\star}}_{1}&\kappa^{{\star}{T}}_{2} & \kappa^{{\star}{\star}}_{2}&  \kappa^{{\star}{T}}_{1} + \kappa^{{\star}{T}}_{2} &
     \kappa^{{\star}{\star}}_{1}+\kappa^{{\star}{\star}}_{2}
\end{bmatrix}
\end{pmatrix}
\end{split}
   \label{eq:jointdistrib}
\end{align}

\noindent where

\begin{align}
    \kappa_{i}=k_{i}(t,t)\\
    \kappa^{\star}_{i}=k_{i}(t,t^{\star})\\
   \kappa^{{\star}{\star}}_{i}=k_{i}(t^{\star},t^{\star}.)
    \label{eq:defkern}
\end{align}

Now, using formula for Gaussian conditional (see Eq. \ref{eq:mvn-cond-s9}-\ref{eq:mvn-cond-s10}), the conditional distribution
of  unknown light curves ($F_{i}, i=1,2$) conditioned on their sum is given as:

\begin{equation}\label{eq:conditional0}
\begin{aligned}
p(F_{i}(t^{\star})|F_{1}(t)+F_{2}(t)) &= G\Big(\mu^{\star}_{i}+\kappa^{{\star}T}_{i}(\kappa_{1}+\kappa_{2})^{-1}(F_{1}(t)+F_{2}(t)-\mu_{1}-\mu_{2}), \\
&\qquad \kappa_{i}^{\star\star}-\kappa_{i}^{{\star}T}(\kappa_{1}+\kappa_{2})^{-1})\kappa_{i}^{\star}\Big)
\end{aligned}
\end{equation}
From above equation,  the log-likelihood of the conditional distribution of $F_{2}^{\star}$ given $F_{tot}$ at the query time instances $t^{\star}$ is easily obtained:
\begin{align}
\begin{split}
\mathcal{L}_{GP} = -\frac{1}{2}\Biggl[ & \sum_{i=1}^{n} \log(2\pi) + \sum_{i=1}^{n} \log(\sigma_{F_2|F_{\text{tot}}(t)}^2(e_i^{\star})) \\
& + \sum_{i=1}^{n} \frac{(F_{\text{tot}}(e_i) - \mu_{F_2|F_{\text{tot}}}(e_i^{\star}))^2}{\sigma_{F_2|F_{\text{tot}}(t)}^2(e_i^{\star})} \Biggr]
\end{split}
\label{eq:loglikegp1}
\end{align}

\noindent where  the number of observed points is $n$,  the variance and mean  of $F_{2}(t^{\star}) $ given $F_{tot}(t)$ are $$\sigma_{F_2|F_{\text{tot}}(t)}^2(e_i^{\star}) = \kappa_{2}^{\star\star} - \kappa_{2}^{\star T} (\kappa_{1} + \kappa_{2})^{-1} \kappa_{2}^{\star}$$ and $$\mu_{F_2|F_{\text{tot}}}(e_i^{\star}) = \mu_{2}^{\star} + \kappa_{2}^{\star T} (\kappa_{1} + \kappa_{2})^{-1} (F_{\text{tot}}(e_{i}) - \mu_{1} - \mu_{2})$$, respectively. This  loglikelihood of the  component
of the signal is integrating (marginalizing) over the possible configurations of the other components.

\section{Posterior sampling statistics} %\appendix
\label{sec:sampstat}
{
Here, we present the posterior distributions of binary parameters inferred through the Bayesian Solver procedure (see Section \ref{sec:Bayes}) for configurations C1-C6 (Table \ref{tab:parameter_combinations}), encompassing a variety of optical time domain and astrometry data observational baselines (see Figs. \ref{fig:Fig1}-\ref{fig:Fig3}).}

\begin{figure}[p]
    \centering
    % First image and subcaption
    \begin{minipage}{0.98\textwidth}
        \centering
        \includegraphics[width=0.6\textwidth]{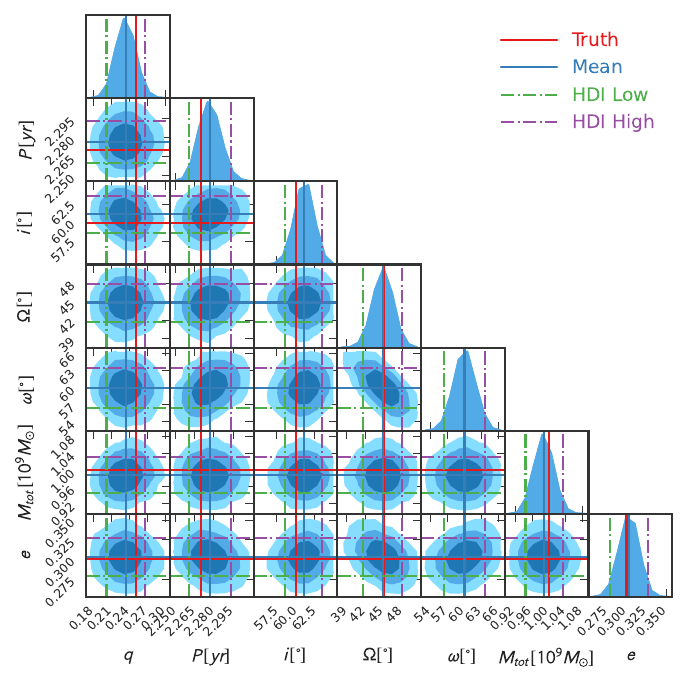}\\
        {(a) $C1$}
    \end{minipage}
    
    % Second image and subcaption
    \begin{minipage}{\textwidth}
        \centering
        \includegraphics[width=0.6\textwidth]{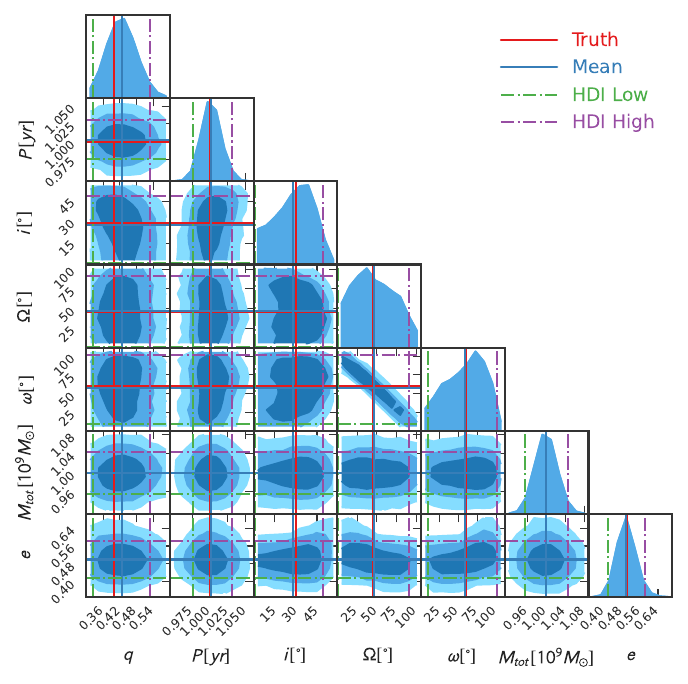}\\
        {(b) $C2$}
    \end{minipage}
    \caption{Posterior probability distributions of the binary  parameters for (a) C1 and (b) C2 configurations, as depicted by the Bayesian models in Fig.  \ref{fig:Fig1}. The intersecting lines on the 2D posterior contours mark truth values of parameters (red), mean posterior (blue), lower (green) and upper (purple) limit of 68\% HDI range. }
    \label{fig:Fig11}
\end{figure}

\newpage
\begin{figure}[p]
    \centering
    % First image and subcaption
    \begin{minipage}{0.98\textwidth}
        \centering
        \includegraphics[width=0.6\textwidth]{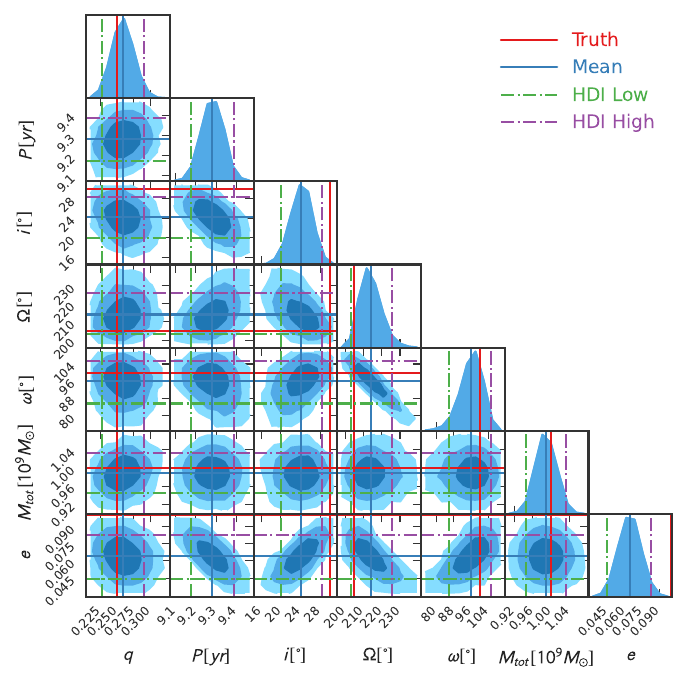}\\
        {(a) $C3$}
    \end{minipage}
    
    % Second image and subcaption
    \begin{minipage}{\textwidth}
        \centering
        \includegraphics[width=0.6\textwidth]{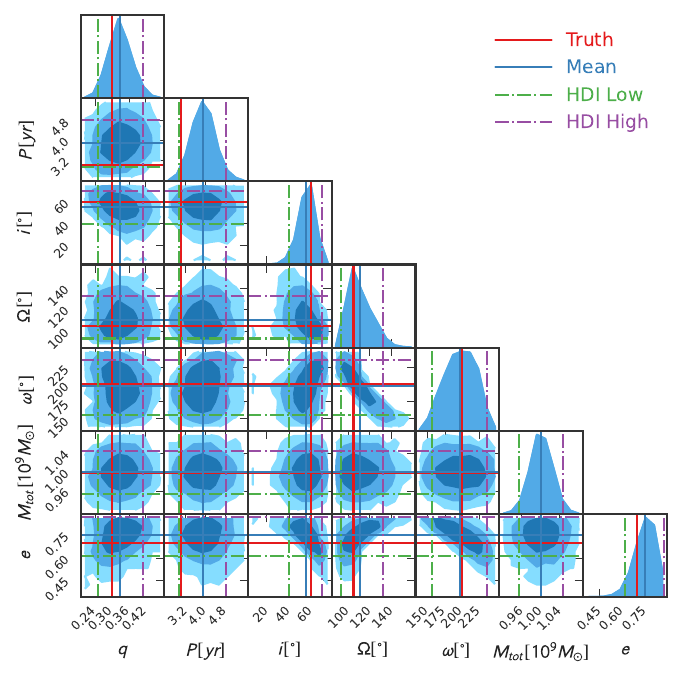}\\
        {(b) $C4$}
    \end{minipage}
    \caption{Bayesian posterior probability distributions of the binary  parameters for configurations (a) $C3$ and (b) $C4$, as depicted by the Bayesian models in Fig. \ref{fig:Fig2}. }
        \label{fig:Fig22}
\end{figure}

\newpage
\begin{figure}[p]
    \centering
    % First image and subcaption
    \begin{minipage}{0.98\textwidth}
        \centering
        \includegraphics[width=0.6\textwidth]{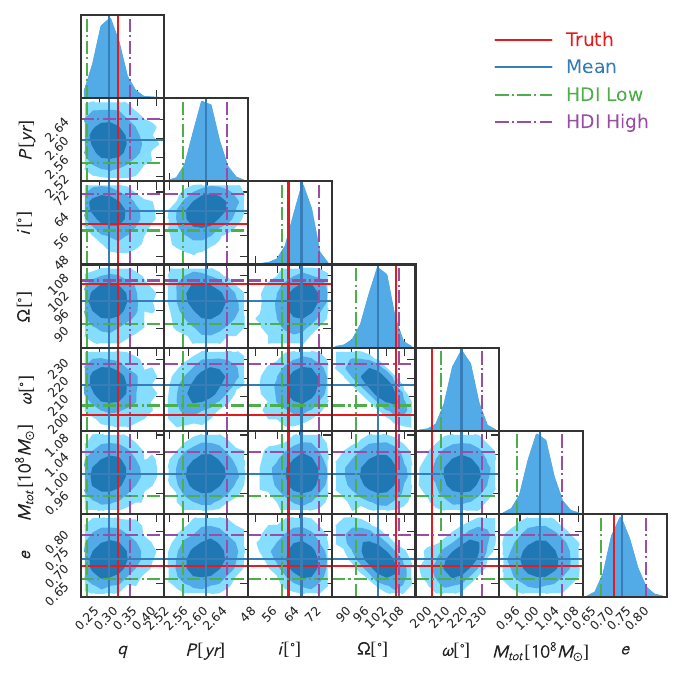}\\
        {(a) $C5$}
    \end{minipage}
    
    % Second image and subcaption
    \begin{minipage}{\textwidth}
        \centering
        \includegraphics[width=0.6\textwidth]{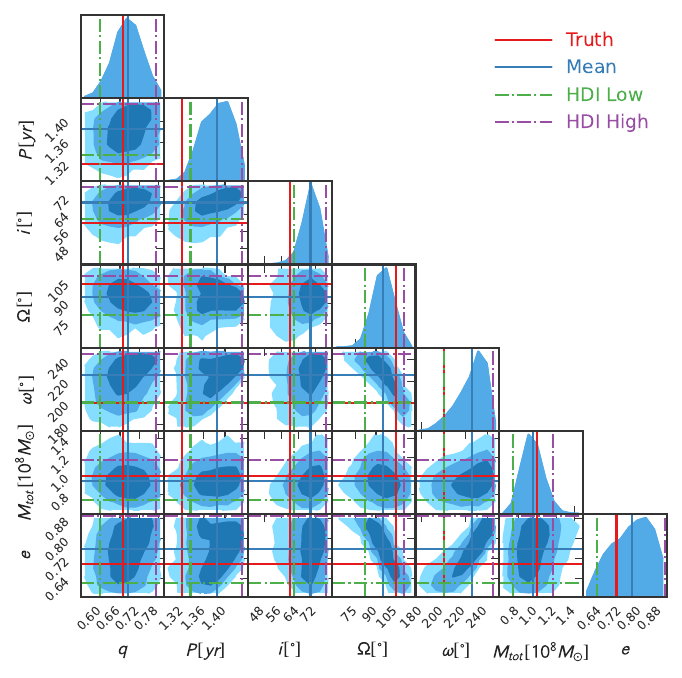}\\
        {(b) $C6$}
    \end{minipage}
    \caption{Bayesian posterior probability distributions of the binary  parameters for  configurations (a) $C5$ and (b) $C6$, as depicted by the Bayesian models in Fig. \ref{fig:Fig3}. }
        \label{fig:Fig33}
\end{figure}

\bibliography{sample631}{}
\bibliographystyle{aasjournal}

%% This command is needed to show the entire author+affiliation list when
%% the collaboration and author truncation commands are used.  It has to
%% go at the end of the manuscript.
%\allauthors

%% Include this line if you are using the \added, \replaced, \deleted
%% commands to see a summary list of all changes at the end of the article.
%\listofchanges

\end{document}